\UseRawInputEncoding
\documentclass[10pt,prl,superscriptaddress,showpacs,
amssymb,amsmath,amsfonts,nofootinbib,floatfix]{revtex4}
\usepackage{graphicx} 
\usepackage{latexsym}
\usepackage{dcolumn} 
\usepackage{bm} 

\usepackage{epsfig}

\usepackage{rotating}
\usepackage{amstext}

\usepackage[dvipsnames]{xcolor}

\newcommand{\gevcsq}{$(\mathrm{GeV/c})^2$}
\newcommand{\qsq}{$Q^2$}

\newcommand{\gep}{$\it G^{p}_{E}$}
\newcommand{\gmp}{$\it G^{p}_{M}$}

\newcommand{\pbo}  {PbWO$_4$}
\newcommand{\moll} {M{\o}ller}
\newcommand{\epreac}   {$ep \to ep$}
\newcommand{\mollreac} {$e^- e^- \to e^- e^- $}
 
\begin{document}
\pagestyle{empty}
\preprint{}
\title{ PRad-II: A New Upgraded High Precision Measurement 
           of the Proton Charge Radius}
\pagenumbering{roman}

\author{A.~Gasparian \footnote{Contact person},\footnote{Spokesperson}}
\affiliation{North Carolina A \& T State University, Greensboro, NC 27424 , USA}
\author{H.~Gao$^\dagger$} 
\affiliation{Duke University, Durham, NC 27708, USA} 
\affiliation{Triangle Universities Nuclear Laboratory, Durham, NC 27708, USA}
\author{D.~Dutta$^\dagger$}
\affiliation{Mississippi State University, Mississippi State, MS 39762, USA}
\author{N.~Liyanage$^\dagger$} 
\affiliation{University of Virginia, Charlottesville, VA 22904, USA}
\author{E.~Pasyuk$^\dagger$}
\affiliation{Thomas Jefferson National Accelerator Facility, Newport News, VA 23606, USA}
\author{D.~W.~Higinbotham$^\dagger$} 
\affiliation{Thomas Jefferson National Accelerator Facility, Newport News, VA 23606, USA}
\author{C.~Peng$^\dagger$}
\affiliation{Argonne National Lab, Lemont, IL 60439, USA}
\author{K.~Gnanvo$^\dagger$}
\affiliation{University of Virginia, Charlottesville, VA 22904, USA}
\author{I.~Akushevich} 
\affiliation{Duke University, Durham, NC 27708, USA} 
\author{A.~Ahmidouch}
\affiliation{North Carolina A \& T State University, Greensboro, NC 27424 , USA}
\author{C. Ayerbe} 
\affiliation{Mississippi State University, Mississippi State, MS 39762, USA}
\author{X.~Bai}
\affiliation{University of Virginia, Charlottesville, VA 22904, USA}
\author{H.~Bhatt}
\affiliation{Mississippi State University, Mississippi State, MS 39762, USA}
\author{D.~Bhetuwal}
\affiliation{Mississippi State University, Mississippi State, MS 39762, USA}
\author{J.~Brock} 
\affiliation{Thomas Jefferson National Accelerator Facility, Newport News, VA 23606, USA}
\author{V.~Burkert}
\affiliation{Thomas Jefferson National Accelerator Facility, Newport News, VA 23606, USA}
\author{D. Byer} 
\affiliation{Duke University, Durham, NC 27708, USA} 
\author{C.~Carlin}
\affiliation{Thomas Jefferson National Accelerator Facility, Newport News, VA 23606, USA}
\author{T.~Chetry}
\affiliation{Mississippi State University, Mississippi State, MS 39762, USA}
\author{E. Christy} 
\affiliation{Hampton University, Hampton, VA 23669, USA }
\author{A.~Deur}
\affiliation{Thomas Jefferson National Accelerator Facility, Newport News, VA 23606, USA}
\author{B.~Devkota}
\affiliation{Mississippi State University, Mississippi State, MS 39762, USA}
\author{J.~Dunne} 
\affiliation{Mississippi State University, Mississippi State, MS 39762, USA}
\author{L.~El-Fassi}
\affiliation{Mississippi State University, Mississippi State, MS 39762, USA}
\author{L.~Gan}
\affiliation{University of North Carolina, Wilmington, NC 28402, USA}
\author{D.~Gaskell}
\affiliation{Thomas Jefferson National Accelerator Facility, Newport News, VA 23606, USA}
\author{Y.~Gotra}
\affiliation{Thomas Jefferson National Accelerator Facility, Newport News, VA 23606, USA}
\author{T. Hague}
\affiliation{North Carolina A \& T State University, Greensboro, NC 27424 , USA}
\author{M.~Jones}
\affiliation{Thomas Jefferson National Accelerator Facility, Newport News, VA 23606, USA}
\author{A.~Karki}
\affiliation{Mississippi State University, Mississippi State, MS 39762, USA}
\author{B. Karki}
\affiliation{Duke University, Durham, NC 27708, USA} 
\author{C.~Keith}
\affiliation{Thomas Jefferson National Accelerator Facility, Newport News, VA 23606, USA}
\author{V. Khachatryan}
\affiliation{Duke University, Durham, NC 27708, USA}
\author{M.~Khandaker}
\affiliation{Energy Systems, Davis, CA 95616}
\author{V.~Kubarovsky}
\affiliation{Thomas Jefferson National Accelerator Facility, Newport News, VA 23606, USA}
\author{I.~Larin} 
\affiliation{University of Massachusetts, Amherst, MA 01003}
\author{D.~Lawrence}
\affiliation{Thomas Jefferson National Accelerator Facility, Newport News, VA 23606, USA}
\author{X. Li}
\affiliation{Duke University, Durham, NC 27708, USA} 
\author{G. Matousek}
\affiliation{Duke University, Durham, NC 27708, USA} 
\author{J.~Maxwell}
\affiliation{Thomas Jefferson National Accelerator Facility, Newport News, VA 23606, USA}
\author{D.~Meekins}
\affiliation{Thomas Jefferson National Accelerator Facility, Newport News, VA 23606, USA}
\author{R.~Miskimen}
\affiliation{University of Massachusetts, Amherst, MA 01003}
\author{S. Mtingwa}
\affiliation{North Carolina A \& T State University, Greensboro, NC 27424 , USA}
\author{V.~Nelyubin}
\affiliation{University of Virginia, Charlottesville, VA 22904, USA}
\author{G. Niculescu}
\affiliation{James Madison University, Harrisonburg, VA 22807, USA}
\author{I. Niculescu}
\affiliation{James Madison University, Harrisonburg, VA 22807, USA}
\author{R.~Pedroni}
\affiliation{North Carolina A \& T State University, Greensboro, NC 27424 , USA}
\author{A.~Shahinyan}
\affiliation{Yerevan Physics Institute, Yerevan Armenia} 
\author{A.P. Smith}
\affiliation{Duke University, Durham, NC 27708, USA} 
\author{S. Srednyak}
\affiliation{Duke University, Durham, NC 27708, USA} 
\author{S.~Stepanyan}
\affiliation{Thomas Jefferson National Accelerator Facility, Newport News, VA 23606, USA}
\author{S.~Taylor}
\affiliation{Thomas Jefferson National Accelerator Facility, Newport News, VA 23606, USA}
\author{E. van Nieuwenhuizen}
\affiliation{Duke University, Durham, NC 27708, USA} 
\author{B.~Wojtsekhowski} 
\affiliation{Thomas Jefferson National Accelerator Facility, Newport News, VA 23606, USA}
\author{W.~Xiong} 
\affiliation{Syracuse University, Syracuse, NY 13244}
\author{B. Yu}
\affiliation{Duke University, Durham, NC 27708, USA} 
\author{Z.~W.~Zhao}
\affiliation{Duke University, Durham, NC 27708, USA} 
\author{J. Zhou}
\affiliation{Duke University, Durham, NC 27708, USA} 
\author{B.~Zihlmann}
\affiliation{Thomas Jefferson National Accelerator Facility, Newport News, VA 23606, USA}
\author{the PRad collaboration.}

\maketitle

\section*{Abstract}
The PRad experiment has credibly demonstrated the advantages of the
calorimetric method in $e-p$ scattering experiments to measure the
proton root-mean-square (RMS) charge radius with high accuracy. The PRad result, within
its experimental uncertainties, is in agreement with the small radius
measured in muonic hydrogen spectroscopy experiments and it was a
critical input in the recent revision of the CODATA recommendation for
the proton charge radius. Consequently, the PRad result is in direct
conflict with all modern electron scattering experiments. Most
importantly, it is 5.8\% smaller than the value from the most precise
electron scattering experiment to date, and this difference is about
three standard deviations given the precision of the PRad experiment.
As the first experiment of its kind, PRad did not reach the highest
precision allowed by the calorimetric technique. Here we propose a new
(and) upgraded experiment -- PRad-II, which will reduce the overall
experimental uncertainties by a factor of 3.8 compared to PRad and
address this as yet unsettled controversy in subatomic physics. In
addition, PRad-II will be the first lepton scattering experiment to
reach the $Q^2$ range of 10$^{-5}$ GeV$^2$ allowing a more accurate
and robust extraction of the proton charge radius.
The muonic hydrogen result with its unprecedented precision (~0.05\%)
determines the CODATA value of the proton charge radius, hence, it is
critical to evaluate possible systematic uncertainties of those
experiments, such as the laser frequency calibration that was raised
in recent review articles. The PRad-II experiment with its projected
total uncertainty of 0.43\% could demonstrate whether there is
any systematic difference between $e-p$ scattering and muonic hydrogen
results. PRad-II will establish a new precision frontier in electron
scattering and open doors for future physics opportunities.
\newpage

\pagenumbering{arabic}

\section{Introduction}
\label{sec:intro}
The proton is the dominant ingredient of visible matter in the Universe. Consequently, determining the proton's basic properties such as its root-mean-square (RMS) charge radius, $r_p$, has attracted tremendous interests in its own right. Accurate knowledge of $r_p$ is essential not only for understanding how strong interactions work in the confinement region, but is also required for precise calculations of the energy levels and transition energies of the hydrogen (H) atom, for example, the Lamb shift. The extended proton charge distribution changes the Lamb shift by as much as $2\%$~\cite{Pohl10} in the case of $\mu$H atoms, where the electron in the H atom is replaced by a "heavier electron", the muon. It also has a major impact on the precise determination of fundamental constants such as the Rydberg constant ($R_{\infty}$)~\cite{Mohr08}. The first principles calculation of $r_p$ in the accepted theory of the strong interaction - Quantum Chromodynamics (QCD), is notoriously challenging analytically and are being carried out by computer simulations,  known as the lattice QCD calculations. Currently, such calculations cannot reach the accuracy demanded by experiments, but are on the cusp of becoming precise enough to be tested experimentally~\cite{lattice}.

Prior to 2010 the two methods used to measure $r_p$ were: (i) \epreac ~elastic scattering measurements, where the slope of the extracted electric form factor (\gep) down to zero 4-momentum transfer squared (\qsq), is proportional to $r_p^2$; and (ii) Lamb shift (spectroscopy) measurements of "regular" H atoms, which, along with state-of-the-art calculations, were used to determine $r_p$. Although, the $e-p$ results can be somewhat less precise than the spectroscopy results, the values of $r_p$ obtained from these two methods~\cite{Bern10,Mohr08} mostly agreed with each other~\cite{CODATA_2012}. New results based on Lamb shift measurements in $\mu$H were reported for the first time in 2010. The Lamb shift in $\mu$H is several million times more sensitive to $r_p$ because the muon is about 200 times closer to the proton than the electron in a H atom. To the surprise of both the nuclear and atomic physics communities, the two $\mu$H results~\cite{Pohl10, Anti13} with their unprecedented, $<$0.1\% precision, were a combined eight-standard deviations smaller than the average value from all previous experiments. This triggered the {\it "proton radius puzzle"}~\cite{carlsonrev}, unleashing intensive experimental and theoretical efforts aimed at resolving this "puzzle". 


The PRad experiment completed in 2016, was the first
high-precision $e-p$ experiment since the emergence of the "puzzle". It was the first electron scattering experiment to utilize a magnetic-spectrometer-free method along with a windowless hydrogen gas target, which overcame several limitations of previous $e-p$ ~experiments and reached unprecedentedly small scattering angles. The PRad result, $r_p =$ 0.831~$\pm$~0.007$_{\rm stat.}$~$\pm$~0.012$_{\rm syst.}$ femtometer, is consistent within uncertainties, with the $\mu$H results and was one of the critical inputs in changing the recent CODATA recommendation for $r_p$~\cite{CODATA_2018}. But, the PRad result is in direct conflict with the world average of all modern $e-p$ results~\cite{CODATA_2014}. For example, it is 5.8\% smaller than the most precise electron scattering experiment to date - the 2010 experiment at Mainz~\cite{Bern10}. In particular, the \gep ~has a systematic difference in the higher \qsq ~range of the PRad experiment.
A new $e-p$ scattering experiment with reduced total uncertainties and exploring the lowest \qsq ~feasible is required to address this as yet unsettled controversy in subatomic physics. As the first experiment of its kind, PRad did not reach the highest precision allowed by the novel magnetic spectrometer-free technique. Therefore, it is timely and incumbent on the collaboration to conduct an upgraded PRad experiment with significantly reduced uncertainties while reaching the lowest scattering angles probed in lepton scattering experiments. 

The $\mu$H result with its unprecedented precision (0.05\%) determines the current average value of $r_p$ ~\cite{CODATA_2018}. Several recent review articles have raised the possibility of additional systematic uncertainties in the $\mu$H results, such as the laser frequency calibration. Therefore it is critical to evaluate all the systematic uncertainties of those experiments. Moreover, among the three most recent H spectroscopy measurements~\cite{CREMA_2017, hspec2018, Eric2019}, two experiments found a small radius~\cite{CREMA_2017,Eric2019} consistent with the $\mu$H results, but they disagree with another one which supports a larger value~\cite{hspec2018}. While the PRad result and those from~\cite{CREMA_2017,Eric2019} are consistent with the $\mu$H results within the experimental uncertainties, the central values from these electron based experiments are all smaller than those from the $\mu$H experiments. These observations have injected a new dimension to the ongoing controversy involving $r_p$ measurements. 
A fundamental difference between the $e-p$ and $\mu-p$ interactions, could be the origin of the discrepancy. However, there are abundant experimental constraints on any such "new physics", and yet models that resolve the puzzle with new force carriers have been proposed~\cite{carlsonrev,jerry2017}. On the other hand, more mundane solutions continue to be explored, for example, the definition of  $r_p$ used in all three major experimental approaches has been rigorously shown to be consistent~\cite{jerry18}. The effect of two-photon exchange on $\mu$H spectroscopy~\cite{jerry13,aa2017} and form factor nonlinearities in $e-p$ scattering~\cite{hill, doug2016, carlson} have also been examined. None of these studies could adequately explain the observations 
and have reinforced the need for additional high-precision measurements of $r_p$, using new experimental techniques with different systematics.    

In summary, the PRad experiment was the first electron scattering experiment to utilize a new technique with completely different systematics compared to all previous magnetic-spectrometer based $e-p$ experiments. The PRad result is consistent with the $\mu$H 
results and consequently it agrees with the recently announced shift in the Rydberg constant~\cite{CODATA_2018}, one of the best-known fundamental constants in physics. The PRad experiment has convincingly demonstrated the validity and advantage of the new calorimetric technique, but further improvements are possible. Here we propose an enhanced version of the PRad experiment with an estimated uncertainty that is a factor of 3.8 smaller than that of the PRad experiment. In addition, it will
be the first lepton scattering experiment to reach the $Q^2$ range of 10$^{-5}$ GeV$^2$ allowing a more accurate and robust extraction of $r_p$. The proposed experiment would make a crucial contribution towards the resolution of the discrepancy between PRad and other modern $e-p$ scattering experiments. The projected total uncertainty of 0.43\% will also be able to address possible systematic difference between $e-p$ and the $\mu$H experiments.
It would then establish a new precision frontier in electron scattering allowing for the exploration of future physics opportunities. 

\section{The PRad experiment}
\label{sec:prad}
\subsection{The novel technique to measure the proton charge radius}
The PRad collaboration at Jefferson Lab developed and performed a new $e-p$ experiment as an independent measurement of $r_p$ to address the "proton radius puzzle". The PRad
experiment, in contrast with previous $e-p$ experiments, was designed to use a magnetic-spectrometer-free, calorimeter based method~\cite{propo}. The innovative design of the PRad experiment enabled three major improvements over previous $e-p$ experiments:
(i)~The large angular acceptance ($0.7^{\circ} - 7.0^{\circ}$) of the hybrid calorimeter (HyCal) allowed for a large \qsq ~coverage spanning two orders of magnitude ($2.1~\times~10^{-4} - 6~\times~10^{-2}$)~\gevcsq, ~in the low \qsq ~range. The single fixed location of \mbox{HyCal} eliminated the multitude of normalization parameters that plague magnetic spectrometer based experiments, where the spectrometer must be physically moved to many different angles to cover the desired range in \qsq. In addition, the PRad experiment reached extreme forward scattering angles down to $0.7^{\circ}$ achieving the lowest \qsq~($2.1~\times~10^{-4}$ \gevcsq~) ~in $e-p$ experiments, ~an order of magnitude lower than previously achieved. Reaching a lower \qsq ~range is critically important since $r_p$ is extracted as the slope of the measured \gep(\qsq) ~at  \qsq ~= 0.  
(ii)~The extracted $e-p$ ~cross sections were normalized to the well
known quantum electrodynamics process - \mollreac ~\moll ~scattering from the atomic electrons ($e-e$)~- which was measured simultaneously with the $e-p$ within the same detector acceptance.
This leads to a significant reduction in the systematic uncertainties of  measuring the $e-p$ ~cross sections.
(iii)~The background generated from the target windows, one of the dominant sources of systematic uncertainty for all previous $e-p$ experiments, is highly suppressed in the PRad experiment.

\begin{figure}[hbt!]
\centerline{\includegraphics[width=0.9\textwidth]{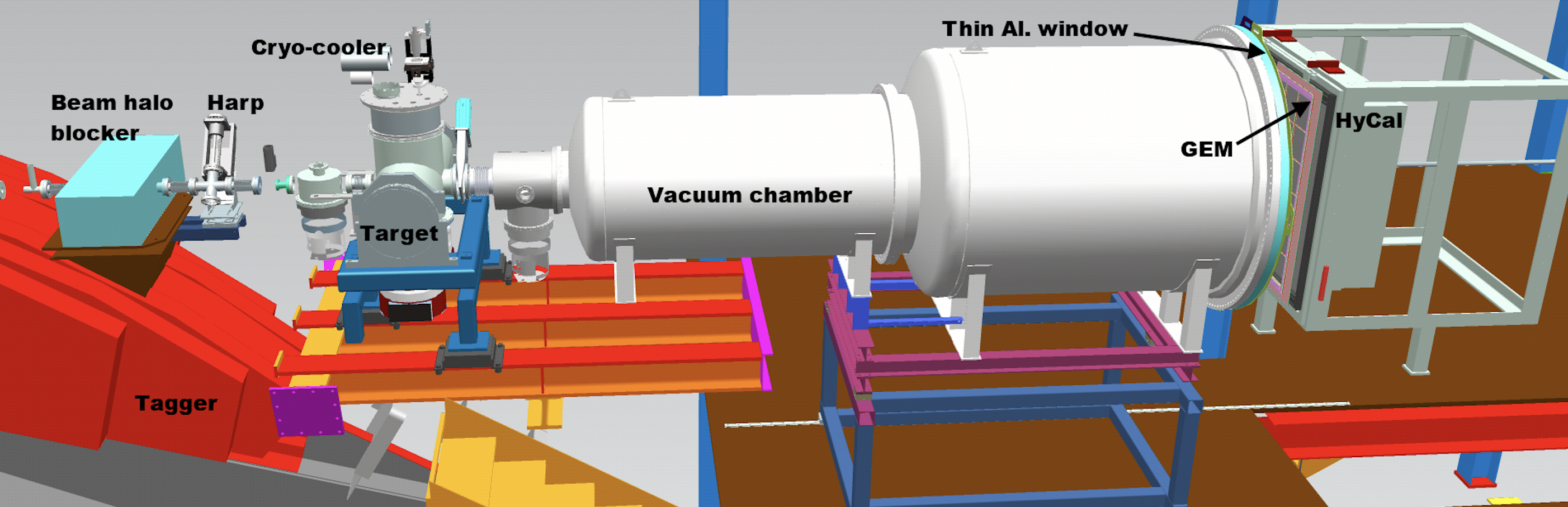}}
\caption{A schematic layout of the PRad experimental setup in \mbox{Hall B} at Jefferson Lab, with the electron beam incident from the left. The key beam line elements are shown along with the window-less hydrogen gas target, the two-segment vacuum chamber and the two detector systems.} 

\label{fig:exp_setup}
\end{figure}
The PRad experimental apparatus consisted of the following four main elements (Figure~\ref{fig:exp_setup}): (i)~a 4~cm long, windowless, cryo-cooled hydrogen (H$_2$) gas flow target with a density of \mbox{$2 \times 10^{18}$} 
atoms/cm$^2$. It eliminated the beam background from the target windows and was the first such target used in $e-p$ experiments;
(ii)~the high resolution, large acceptance \mbox{HyCal} electromagnetic calorimeter~\cite{HyCal}. The complete azimuthal coverage of \mbox{HyCal} for the forward scattering angles allowed simultaneous detection of the pair of electrons from $e-e$ ~scattering, for the first time in these types of measurements; 
(iii)~one plane made of two high resolution $X-Y$ gas electron multiplier (GEM)  coordinate detectors located in front of \mbox{HyCal}; and
(iv)~a two-section vacuum chamber spanning the 5.5~m distance from the target to the detectors.

\begin{figure}[hbt!]
\centerline{\includegraphics[width=0.5\textwidth]{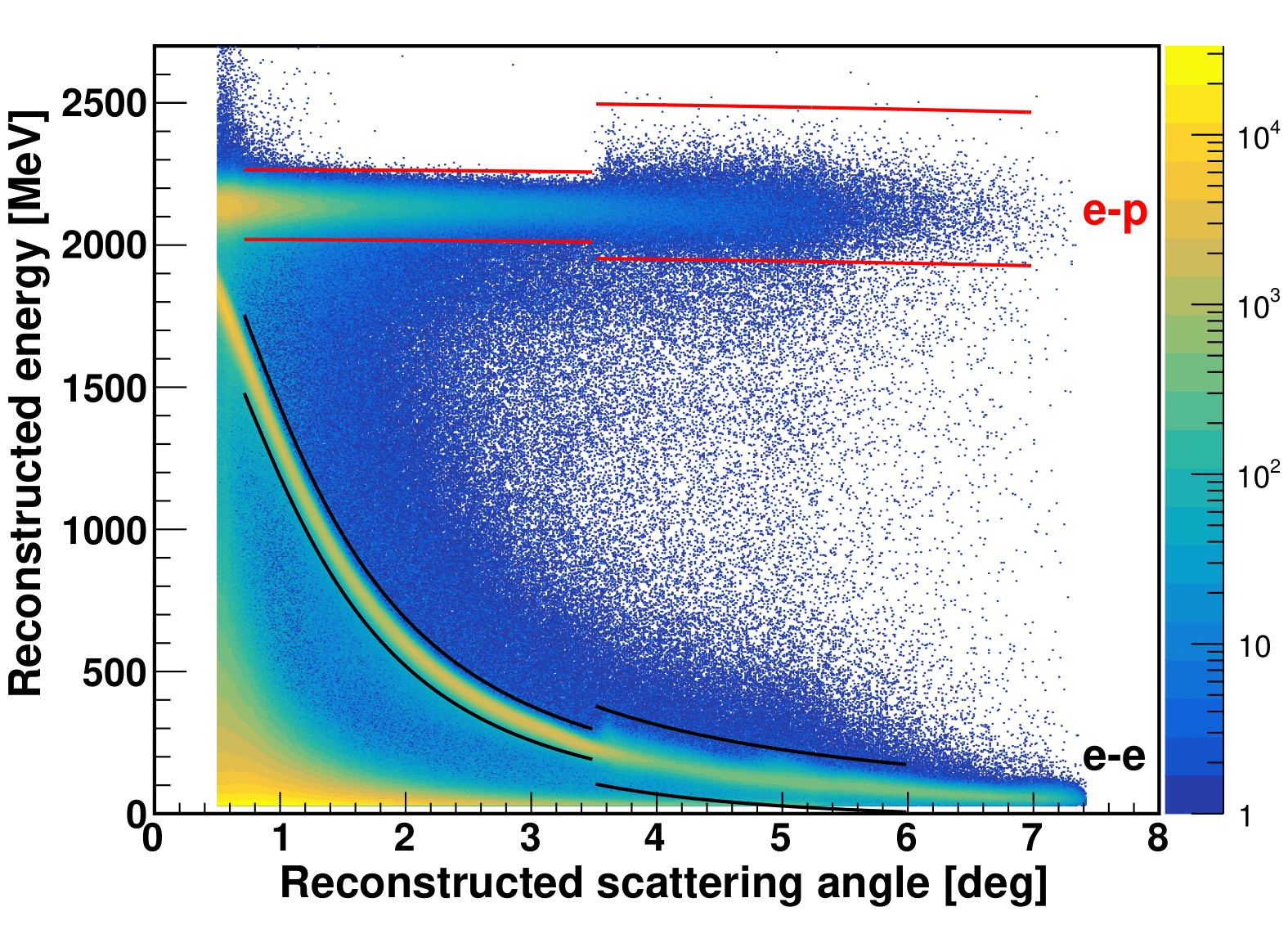}\hspace{2ex}\includegraphics[width=0.5\textwidth]{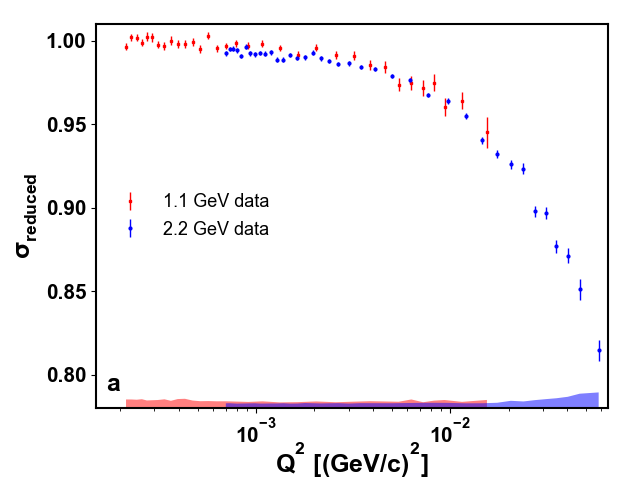}}
\caption{The reconstructed energy vs angle for $e-p$ and $e-e$ events for the electron beam energy of 2.2 GeV. The red and black lines indicate the event selection for $e-p$ and $e-e$, respectively. The angles $\leq 3.5^{\circ}$ are covered by the PbWO$_4$ crystals and the rest by the Pb-glass part of \mbox{HyCal}.} 
\label{fig:recon2}
\end{figure}

The PRad experiment was performed in \mbox{Hall B} at Jefferson Lab in May-June of 2016, using 1.1~GeV and 2.2~GeV electron beams. The standard \mbox{Hall B} beam line, designed for low beam currents (0.1-50~nA), was used in this experiment. The incident electrons that scattered off the target protons and the \moll ~electron pairs, were detected in the GEM and \mbox{HyCal} detectors. The energy and position of the detected electron(s) was measured by \mbox{HyCal}, and the transverse ($X-Y$) position was measured by the GEM detector, which was used to assign the \qsq ~for each detected event. The GEM detector, with a position resolution of 72~$\mu$m, ~improved the accuracy of \qsq ~determination. Furthermore, the GEM detector suppressed the contamination from photons generated  in the target and other beam line materials; the \mbox{HyCal} is equally sensitive to electrons and photons while the GEM is mostly insensitive to neutral particles. The GEM detector also helped suppress the position dependent irregularities in the response of the electromagnetic calorimeter. A plot of the reconstructed energy versus the reconstructed angle for $e-p$ and $e-e$ ~events is shown in Figure~\ref{fig:recon2}~for the 2.2 GeV beam energy.

The background was measured periodically with an empty target cell. To mimic the residual gas in the beam line, H$_2$ gas at very low pressure was allowed in the target chamber during the empty target runs. The charge normalized $e-p$ and M{\o}ller yields from the empty target cell  were used to effectively subtract the background contributions. The beam current was measured with the Hall-B Faraday cup with an uncertainty of $<$~0.1\%~\cite{clasnim}.

A comprehensive Monte Carlo simulation of the PRad setup was developed using the Geant4 toolkit~\cite{geant4}. 
The simulation consists of two separate event generators built for the $e-p$ and $e-e$  processes~\cite{Akushevich:2015toa, Gramolin:2014pva}. Inelastic $e-p$ scattering events were also included in the  simulation using a fit~\cite{Christy:2007ve} to the $e-p$ inelastic world data. The simulation included signal digitization and photon propagation which were critical for the precise reconstruction of the position and energy of each event in the \mbox{HyCal}.

\begin{figure}[hbt!]
\centerline{\includegraphics[width=0.5\textwidth]{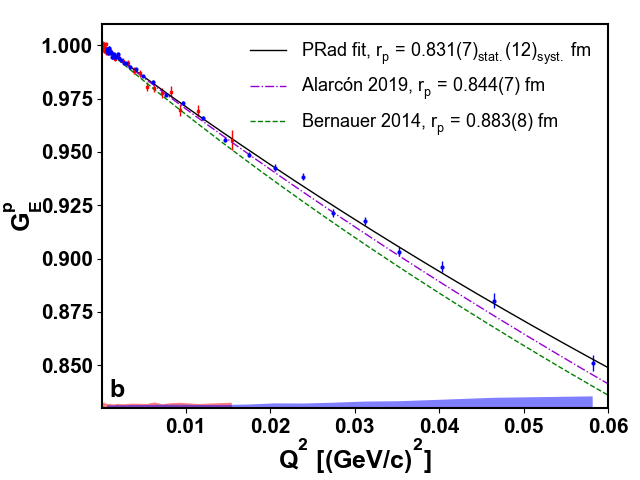}
\includegraphics[width=0.5\textwidth]{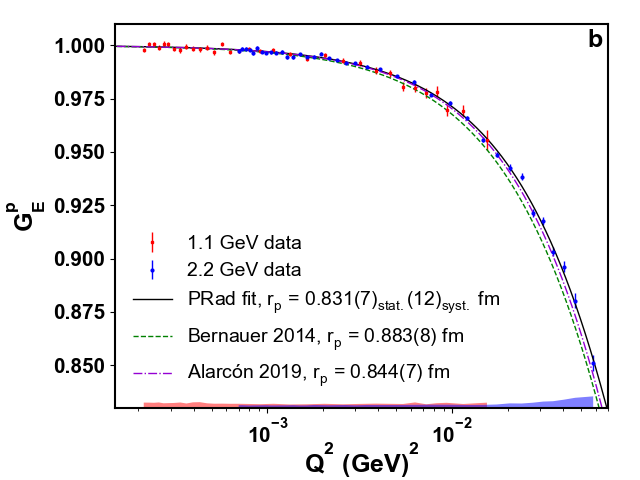}}
\caption{\textbf{(a)} The reduced cross section ($\sigma_{\mbox{reduced}} = \left(\frac{d\sigma}{d\Omega}\right)_{e-p}/\left[\left(\frac{d\sigma}{d\Omega}\right)_{\mbox{point-like}} \left(\frac{4M_p^2E'}{(4M_p^2+Q^2)E}\right)\right]$, where $E$ is the electron beam energy, $E'$ is the energy of the scattered electron and $M_p$ is the mass of the proton), for the PRad $e-p$ data. Dividing out the kinematic factor inside the parentheses, the reduced cross section is a linear  combination of the electromagnetic form factors squared. The systematic uncertainties are shown as bands. 
\textbf{(b)} The  $G^p_E$ as a function of $Q^{2}$. The data points are normalized with the $n_{1}$ and $n_{2}$ parameters, for the 1.1 GeV and 2.2 GeV data separately. Statistical uncertainties are shown as error bars. Systematic uncertainties are shown as bands, for 1.1 GeV (red) and 2.2 GeV (blue). The solid black curve shows the $G_{E}(Q^{2})$ from the fit to the function given by Eq.~\ref{eq:rational}. Also shown are the fit from a previous $e-p$ experiment~\cite{Bern10} for $r_p =$ 0.883(8) fm (green) and the calculation of Alarcon {\it et al.}~\cite{alarcon} for $r_p =$ 0.844(7) fm (purple).}
\label{fig:GE}
\end{figure}
The $e-p$ cross sections were obtained by comparing the simulated and measured $e-p$ ~yield relative to the simulated and measured $e-e$ yield. The extracted reduced cross section  is shown in Figure~\ref{fig:GE}~(a). The $e-p$ elastic cross section is related to \gep ~and the proton magnetic form factor (\gmp) ~as per the Rosenbluth formula~\cite{propo}. In the very low $Q^{2}$ region covered by the PRad experiment, the cross section is dominated by the contribution from \gep. ~Thus, the uncertainty introduced from \gmp ~is negligible. In fact, when using a wide variety of parameterizations for $G^p_{M}$~\cite{Kelly:2004hm, Venkat:2010by, Bern10, Higinbotham:2019jzd}, the extracted $G^p_{E}$ varies by $\sim $~0.2\% at \qsq~$= 0.06$~\gevcsq, the largest \qsq ~accessed by the PRad experiment, and $<$~0.01\% in the \qsq $<~0.01$~\gevcsq~region. The largest variation in $r_p$ arising from the choice of $G^p_{M}$ parametrization is $0.001$~fm. The $G^p_{E}(Q^2)$ extracted from our data is shown in Figure~\ref{fig:GE}~(b), where the Kelly parametrization for $G^p_{M}$~\cite{Kelly:2004hm} was used.

\subsection{The results}

The slope of $G^p_E(Q^{2})$ ~as $Q^{2}~\rightarrow$~0 is proportional to $r_p^2$. A common practice is to fit $G^p_E(Q^{2})$ to a functional form and to obtain $r_{p}$ by extrapolating to $Q^{2}=$~0. However, each functional form truncates the higher-order moments of $G^p_E(Q^2)$ differently and introduces a model dependence which can bias the determination of $r_{p}$.
It is critical to choose a robust functional form that is most likely to yield an unbiased estimation of $r_{p}$ given the uncertainties in the data, and test the chosen functional form over a broad range of parameterizations of $G^p_E(Q^{2})$~\cite{Yan:2018bez}. To simultaneously minimize the possible bias in the radius extraction and the total uncertainty, various functional forms were examined for their robustness in reproducing an input $r_p$ used to generate a mock data set 
 that had the same statistical uncertainty as the PRad data. The robustness quantified as the root mean square error (RMSE) is defined as ${\mbox{RMSE}} = \sqrt{(\delta R)^2 + \sigma^2}$, where $\delta R$ is the bias or the difference between the input and extracted radius and $\sigma$ is the statistical variation of the fit to the mock data~\cite{Yan:2018bez}. These studies show~\cite{Yan:2018bez} ~that consistent results with the least uncertainties can be achieved when using the multi-parameter Rational-function  (referred to as Rational (1,1)): 
\begin{equation} \label{eq:rational}
f(Q^{2}) = nG_{E}(Q^{2}) = n\frac{1+p_{1}Q^{2}}{1+p_{2}Q^{2}},
\end{equation}
where $n$ is the floating normalization parameter, and the charge radius is given by $r_p = \sqrt{6(p_{2} - p_{1})}$.
The $G^p_E(Q^{2})$ extracted from the 1.1~GeV and 2.2~GeV data were fitted simultaneously using the Rational (1,1) function. Independent normalization parameters $n_{1}$ and $n_{2}$ were assigned for 1.1 and 2.2 GeV data respectively, to allow for differences in normalization uncertainties, but the $Q^{2}$ dependence was identical.
The parameters obtained from fits to the Rational (1,1) function are: $n_{1} = 1.0002 \pm 0.0002_{\rm{stat.}} \pm 0.0020_{\rm{syst.}}$, $n_{2} = 0.9983 \pm 0.0002_{\rm{stat.}} \pm 0.0013_{\rm{syst.}}$, and $r_{p} = 0.831 \pm 0.007_{\rm{stat.}} \pm 0.012_{\rm{syst.}}~\rm{fm}$. 
The Rational (1,1) function describes the data very well, with a reduced $\chi^{2}$ of 1.3 when considering only the statistical uncertainty. 

To determine the systematic uncertainty in $r_p$, a Monte Carlo technique was used to randomly smear the cross section and $G_{E}(Q^2)$ data points for each known source of systematic uncertainty. The $r_p$ was extracted from the smeared data and the process is repeated 100,000 times. The RMS of the resulting distribution of $r_{p}$ is recorded as the systematic uncertainty. The dominant systematic uncertainties of $r_{p}$ are the $Q^{2}$ dependent ones which primarily affect the lowest-$Q^{2}$ data: the M{\o}ller radiative corrections, the background subtraction for the 1.1~GeV data, and event selection. The uncertainty of $r_p$ arising from the finite \qsq ~range and the extrapolation to \qsq ~=~0, was investigated by varying the \qsq ~range of the mock data set as part of the robustness study of the Rational (1,1) function~\cite{Yan:2018bez}.  This uncertainty was found to be much smaller than the relative statistical uncertainty of $0.8$\%. The total systematic relative uncertainty on $r_p$ was found to be 1.4\%. 

\begin{figure}[hbt!]
\centerline{\includegraphics[width=1.0\textwidth]{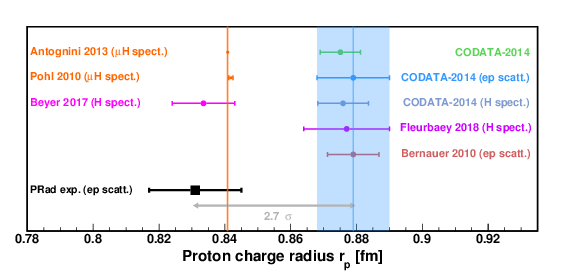}}  
\caption{The proton charge radius. The $r_p$ extracted from the PRad data, shown along with the other measurements of $r_p$ since 2010 and the CODATA recommended values.The PRad result is 2.7-$\sigma$ smaller than the CODATA recommended value for $e-p$ experiments~\cite{CODATA_2014}.}
\label{fig:proton_radius}
\end{figure}

The $r_{p}$ obtained using the Rational (1,1) function is shown in Figure~\ref{fig:proton_radius}, with statistical and systematic uncertainties summed in quadrature. Our result obtained from \qsq ~down to an unprecedented $2.1 \times 10^{-4}$ (GeV/c)$^2$, is about 3-standard deviations smaller than the previous high-precision electron scattering measurement~\cite{Bern10}, which was limited to higher \qsq ~($> 0.004$ (GeV/c)$^2$). 
On the other hand, our result is consistent with the $\mu$H Lamb shift measurements\cite{Pohl10, Anti13}, and also the recent 2S-4P transition frequency measurement using ordinary H atoms~\cite{CREMA_2017}. Given that the lowest \qsq ~reached in the PRad experiment 
is an order of magnitude lower than the previous $e-p$ experiments, and the careful control of systematic effects, our result indicates that the proton is indeed smaller than the previously accepted value from $e-p$ measurements. Our result does not support any fundamental difference between the $e-p$ and $\mu-p$ interactions and is consistent with the shift in the Rydberg constant announced by CODATA~\cite{CODATA_2018}.

In summary, the PRad experiment is the first $e-p$ experiment to cover a two orders of magnitude span of \qsq, ~in one setting. The experiment also exploited the simultaneous detection of $e-p$ and $e-e$ scattering to achieve superior control of systematic uncertainties, which were by design different from previous $e-p$ experiments. Further, the extraction of $r_p$ by employing  functional forms with validated robustness
is another strength of this result. Our result introduces a large discrepancy  with contemporary
  precision $e-p$ experiments. On the other hand, the  results also imply that there is consistency between proton charge radii obtained from $e-p$ scattering 
on regular hydrogen and spectroscopy of muonic hydrogen~\cite{Pohl10,Anti13} and that the value of $r_p$ is consistent with the recently updated CODATA value~\cite{CODATA_2018}. The PRad experiment demonstrates the clear advantages of the calorimeter based method for extracting $r_p$ from $e-p$ experiments and points to further possible improvements in the accuracy of this method. It also validates the recently announced shift in the Rydberg constant~\cite{CODATA_2018}, which has profound consequences, given that the Rydberg constant is one of the most precisely known constants of physics. 


\section{PRad-II: Going beyond-the-state-of-the-art}
\label{sec:apparatus}
Based on the experience gained from the PRad experiment, there are a number of improvements one can make in an upgraded experiment. The low-hanging fruits amongst the possible improvements are additional beamtime to reduce the statistical uncertainty and a better beamline vacuum upstream of the target, to help reduce the small angle experimental background. The rest of the improvements are related to reducing the key systematic uncertainties that dominated the PRad experiment, namely i) the precision of the efficiency determination of the GEM based coordinate detector, ii) the need to cover a wider range of \qsq ~than PRad, by reaching the lowest \qsq ~accessed by lepton scattering experiments  iii) the non-linear detector response of the Pb-Glass portion of the HyCal calorimeter and iv) the subtraction of background associated with the beam line. Lastly, improved radiative correction calculations will further improve the precision of the proton radius determination from the PRad and the upgraded PRad-II experiment. The improved radiative corrections will be discussed later in the proposal in section~\ref{sec:newrad}.

\subsection{New tracking capabilities}
\label{sec:new_tracker}
The precision of the GEM detector efficiency contributed indirectly to the systematic uncertainty of the PRad experiment. A precise measurement of the GEM detector efficiency (at the level of 0.1\%) allows the integrated M{\o}ller method to be used over the entire angular acceptance of the experiment. The uncertainties associated with M{\o}ller counts used in this method are normalization type uncertainties and thus, do not contribute to the systematic uncertainty of extracting $r_{p}$. However, this method relies on a correction for the inefficiency of the GEM detector. As can be seen in fig.~\ref{fig:gem_effres}-right,  the presence of the spacer grids (which are  used to keep the GEM foils apart from each other) in the PRad GEM detectors caused narrow regions of lower efficiency along the spacers. While these efficiencies were measured relative to HyCal and corrected in data analysis, the relatively poor position resolution of the HyCal led to larger uncertainties in the locations of these low efficiency  areas of the GEM detectors. This resulted in systematic uncertainties as large as 0.5\% in the forward scattering angular region. These larger systematic uncertainties precluded the integrated M{\o}ller method from being applied in the forward angle region. Instead, the PRad result relied on the bin-by-bin method for the forward angle region. While the bin-by-bin method is excellent in canceling the effect of the GEM detector inefficiency, it introduces $Q^{2}$-dependent systematic uncertainties due to the angular dependence of M{\o}ller scattering with contributions from M{\o}ller radiative correction, M{\o}ller event selection, beam energy and acceptance. Higher precision in the determination of the GEM efficiency would allow for the use of the integrated M{\o}ller method over the full experimental acceptance eliminating these $Q^{2}$-dependent systematic uncertainties.

Using new GEM detectors with no spacer grids significantly reduces the efficiency fluctuations across the active area. Furthermore, a high precision measurement of the GEM detector efficiency profile can be achieved by adding a second GEM detector plane. In this case, each GEM plane can be calibrated with respect to the other GEM plane instead of relying on the HyCal, minimizing the influence of the HyCal position resolution. It will also help reduce various backgrounds such as, cosmic backgrounds and the high-energy photon background that have an impact on the determination of the GEM efficiency.  In addition, the tracking capability afforded by the pair of separated GEM planes will allow measurements of the interaction $z-$vertex. This can be used to eliminate various beam-line backgrounds, such as those generated from the upstream beam halo blocker. The uncertainty due to the subtraction of the beamline background, at forward angles, is one of the dominant uncertainties of PRad. Therefore, 
the addition of the second GEM detector plane will reduce the systematic uncertainty contributed by two  dominant sources of uncertainties.

The tracking capabilities of PRad-II will be enhanced  significantly compared to PRad  by replacing the original GEM layer with a new GEM-type coordinator detector with no spacer grid and with the addition of a second GEM layer, 40 cm upstream of the first  GEM location  next to HyCal. These  two new tracking layers will be built by the UVa group. The outer dimensions and readout parameters of these new layers will be similar to the original PRad GEM layer; with an  active area of  123 cm $\times$ 110 cm composed of  two side by side  detectors,  each with an  active area of 123 cm $\times$ 55 cm, arranged so that there is a narrow overlap area in the middle. 
These two new tracking layers  will be based on the novel  $\mu$RWELL technology. The biggest advantage of using this new technology for the PRad-II tracking layers is that it would allow each detector module to be built without a spacer grid.  The presence of the spacer grid in the original GEM detector caused narrow regions of lower efficiency along the spacers.  Having two spacer-less  layers  will eliminate the regions of low efficiency.  Furthermore, having two  layers  allows for highly accurate  determination of efficiency profile for the entire GEM area; i.e much smaller inefficiency corrections to make  and the inefficiency corrections determined with much higher accuracy. 

\begin{figure}[!ht]
\centerline{
\includegraphics[width=1.0\textwidth, angle = 0]{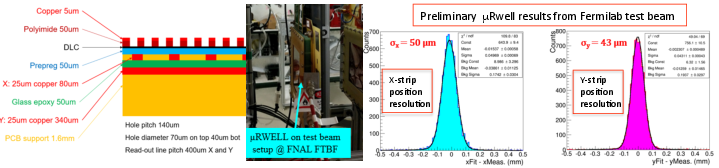}}
\caption{{\bf{ $\mu$RWELL prototype with 2D readout}}: (\textit {left}:) Cross section of the prototype; (\textit {center}:)  Prototype installed in test beam area at Fermilab (June-July 2018); (\textit {right}): Preliminary results of spatial resolution performances  of  the $\mu$RWELL prototype with 2D  X-Y strip readout layer.}
\label{fig:urwell}

\end{figure}

$\mu$RWELL is a single-stage amplification Micro Pattern Gaseous Detector (MPGD) that is a derivative of the  GEM technology. It features a single kapton foil with GEM-like conical holes that are closed off at the bottom by gluing the kapton foil to a readout structure to form a microscopic {\it well} structure. A cross section of a $\mu$RWELL detector is shown in Fig.~\ref{fig:urwell}-left. The technology shares similar performances with a GEM detector in term of  rate capability and  position resolution but presents the advantages of flexibility, no need for spacers  and lower production cost that makes it the ideal candidate for large detectors.  The UVa group built   a 10 $\times$ 10 cm$^2$ $\mu$RWELL prototype detector that was tested  with   cosmic-rays  at the UVa Detector Lab as well as in test beam at Fermilab (June-July 2018). Preliminary results from the test beam data, shown shown in Fig.~\ref{fig:urwell}-right,  are very encouraging with spatial resolution performances superior to those of standard Triple-GEM detectors of similar dimensions. The UVa group  plans to continue basic R\&D studies of the MPGD technology and build large area flat $\mu$RWELL structures for the  PRad-II setup.  

\subsection{Enhancing the Q$^2$ coverage}
\begin{figure}[hbt!]
\centerline{\includegraphics[width=0.525\textwidth]{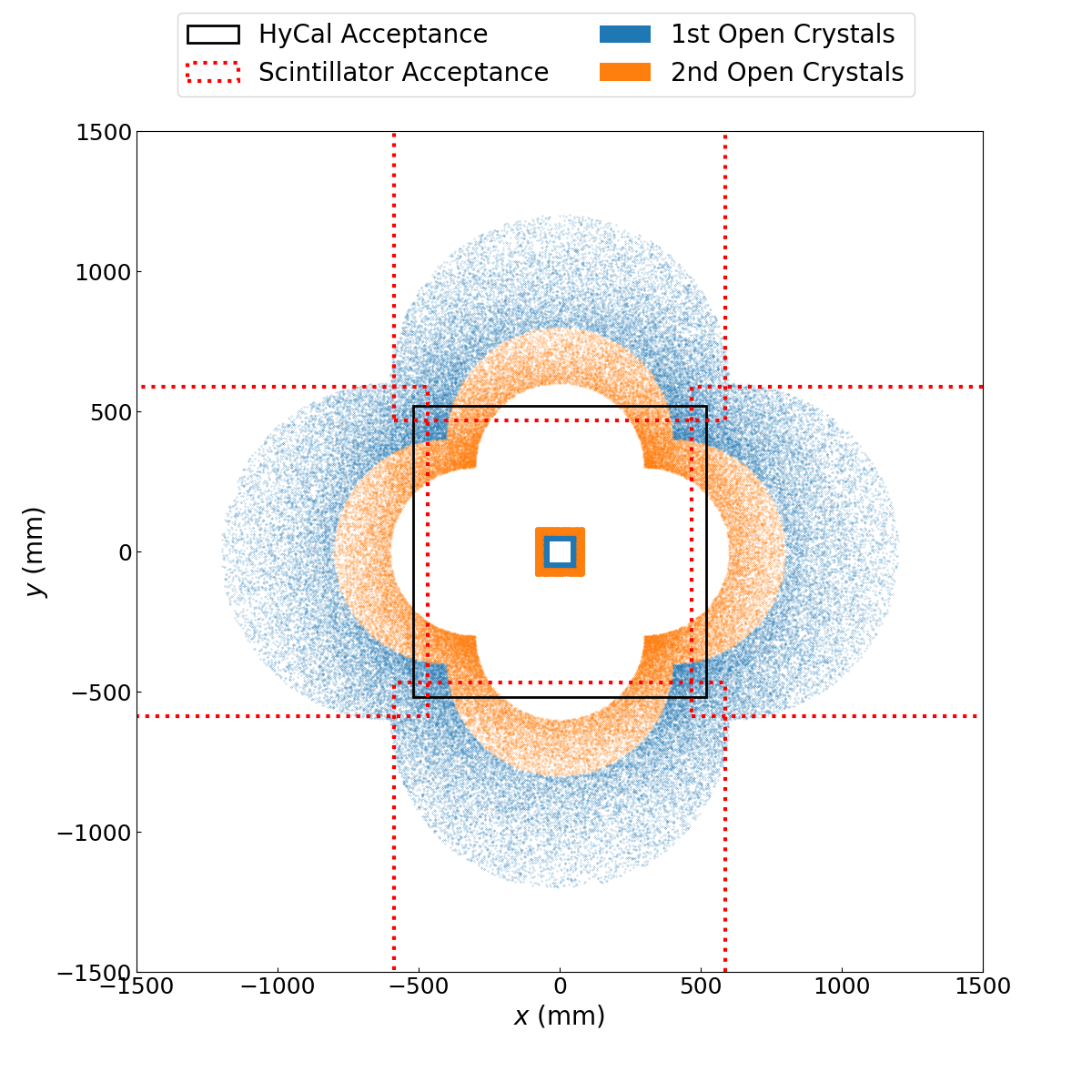}\hspace{5ex}\includegraphics[width=0.4\textwidth]{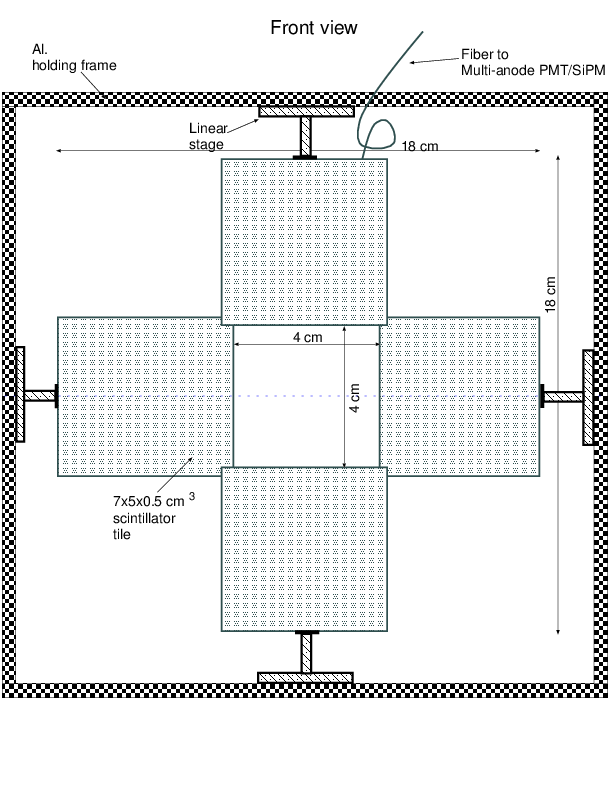}}
\caption{a) The distribution of a \moll ~electron on the scintillator detector, when the other \moll ~electron is detected in the two inner-most layers of the HyCal. The red lines show the outline of scintillator tiles. b) A schematic of the scintillator detector to detect these \moll ~electrons  which  help can be used to separate them from the $e-p$ electrons in the 0.5-0.8 deg range of scattering angles.}
\label{fig:prad2_scin}
\end{figure}

PRad-II will cover a significantly larger range in Q$^2$ compared to the PRad experiment. It will reach an unprecedented low Q$^2$ of $\sim$ 10$^{-5}$ GeV$^2$ while simultaneously covering up to Q$^2$ = 6$\times$ 10$^{-2}$ GeV$^2$. The entire range will be covered in a single fixed experimental setup, just as in PRad, using 3 different beam energies of 0.7, 1.4 and 2.1 GeV. In order to reach the lowest scattering angles of up to 0.5 deg a new rectangular cross shaped scintillator detector will be placed 25~cm from the target center. The scintillator detector covers the angular range beyond the largest angles reached by HyCal. This detector will be used to separate the elastic $e-p$ events from \moll ~scattering events down to scattering angles as low as 0.5 degree thus reaching \qsq ~of $\sim$ 10$^{-5}$ GeV$^2$. The distribution of the \moll ~electrons where the second  scattered electrons is detected in the two inner-most layers of PbWO$_4$ crystals in the HyCal is shown in Fig.~\ref{fig:prad2_scin}(a). For most of the \moll~electrons incident at the two inner-most layers of HyCal, the second \moll ~electron falls outside the HyCal acceptance. By detecting this  second M{\o}ller electron in a scintillator detector placed at $z=$ 25 cm from the center of the target, as shown in Fig.~\ref{fig:prad2_scin}(b), the $e-p$ electrons can be distinguished from the \moll ~electrons at the scattering angles between 0.5-0.8 degrees helping reach lower \qsq ~compared to the PRad experiment. \\  

The detector system will include 4-linear stages such that each scintillator tile can be moved individually in x/y direction enabling them to intersect with the electrons detected in the HyCal. This feature will be used to calibrated the detector and determine its efficiency at the level of 0.1\%. An alpha source based monitoring system will be included on each scintillator tile to help monitor the efficiency continuously during the experiment.\\

\subsection{An all PbWO$_4$ Calorimeter with flash-ADC based readout}
The non-linear behaviors in the energy response of the Pb-glass shower detectors are usually few time
 larger than PbWO$_{4}$ detectors and also much more non-uniform. In addition, their energy resolution is about 2.5 times worse than that of the PbWO$_{4}$ detectors, which increases the inelastic $e-p$ contribution to the elastic $e-p$ yield. Even though the  contributions of these factors  to the $r_{p}$ systematic uncertainty are not as large as those from the M{\o}ller, their contributions to the cross section and $G_{E}^{p}$ are much larger and primarily affect the high $Q^{2}$ data. The only way to reduce this uncertainty is to replace the Pb-glass detectors with PbWO$_{4}$ detectors. This will suppress the inelastic $e-p$ contribution to less than 0.1\% for the entire $Q^{2}$ range, compared to the maximum 2\% in the case of PRad. And it will suppress the $Q^{2}$-dependent systematic uncertainties due to differences in the detector properties between the PbWO$_{4}$ and Pb-glass detectors. Further, converting
 the calorimeter readout electronics from a FASTBUS based system to a flash analog to digital converter based setup would dramatically improve the uncertainty due to detector gain and pedestal stability. The flash-ADC readout system not only allows one to measure the pedestal event-by-event, but also provides excellent timing information and digital trigger information, which allows the rejection of various accidental events and improved trigger efficiency.

PRad-II will use an upgraded HyCal calorimeter which will be an all PbWO$_4$ Calorimeter rather than the Hybrid version used in PRad. The lead-glass modules of HyCal will be replaced with new PbWO$_4$ crystals. This will significantly improve the uniformity of the electron detection over the entire experimental acceptance. Such uniformity of the detector package is critical for the precise and robust extraction of $r_p$. Moreover, the readout electronics will be converted from a FASTBUS based system used during PRad to an all flash-ADC based system which is expected to provide a seven fold improvement in the DAQ speed. A faster DAQ will allow us to collect an order of magnitude more statistics within a reasonable amount of beamtime. Note that the projected uncertainties can still be achieved with the current hybrid calorimeter.

\subsection{Beamline enhancements}
The window on the Hall-B tagger is being replaced with an aluminum windows; this upgrade  is expected to result in a significant improvement in the beamline vacuum, particularly upstream of the target. This will help reduce one of the key sources of background observed during the PRad experiment. Further, a new beam halo blocker will be placed upstream of the Hall-B tagger magnet. This will further reduce the beam-line background critical for accessing the lowest angular range and hence the lowest Q$^2$ range in the experiment. 


\subsection{Other desirable upgrades - Windowless target}
The proposed reduction in the total uncertainties of PRad-II does not rely on the upgrade discussed below but it is
nonetheless desirable for performing the best possible experiment.

The factor of 3.8 improvement expected in the PRad-II experiment does not require any further improvements to the
windowless gas flow target used during PRad. The projected precision and all of the experimental goals can be achieved with the existing PRad target. However, because of recent technological advances, the experiment could benefit from a liquid-drop hydrogen target with a laser based gating. The PRad target is a window-less target which produces two gas plumes escaping from the two ends of the target cell. The effect of these plumes cannot be completely subtracted using the "empty" target runs, where the target chamber is filled with gas at the same flow rate as the gas filling the target cell during the "full" target run. The effect of the plumes was estimated using the PRad target profile simulations. It was difficult to further reduce the systematic uncertainty contributed by the plumes, as the gas profile simulation at low densities is highly non-trivial. A liquid-drop target with an adequate gating mechanism, on the other hand, will be effectively a point-like target, and it should minimize systematic uncertainties associated with the extended target effects, including these plumes. Such a target is being investigated by the JLab target group. However, all the estimates in this proposal are based on the existing PRad gas flow target.


\section{The new proposed experiment}
\label{new_expt}

\subsection{Introduction}
The proposed PRad-II experiment plans to reuse the PRad setup (shown in Fig.~\ref{fig:exp_setup} but with improvements to the range of $Q^2$ covered, an additional GEM detector plane, and an improved high efficiency \pbo ~crystal electromagnetic calorimeter together with a new fADC based readout system for the calorimeter. Just as in the PRad exeriment the scattered electrons from $ep$ elastic and M{\o}ller scatterings will be detected simultaneously with high precision. As demonstrated by the PRad experiment a windowless target cell has a definitive advantage over closed cell targets in minimizing one of the primary sources of background. 

A small scintillator detector placed 25 cm from the target cell will help distinguish between \moll~electrons and and $ep$ elastic electrons at the lowest angles covered in this experiment (0.5-0.7 deg), allowing access to an unprecedented low $Q^2$ of 10$^{-5}$ GeV$^2$.\\

\begin{figure}[!hbt]
\vskip 0.35truecm
\centerline{
\includegraphics[width=0.9\textwidth]{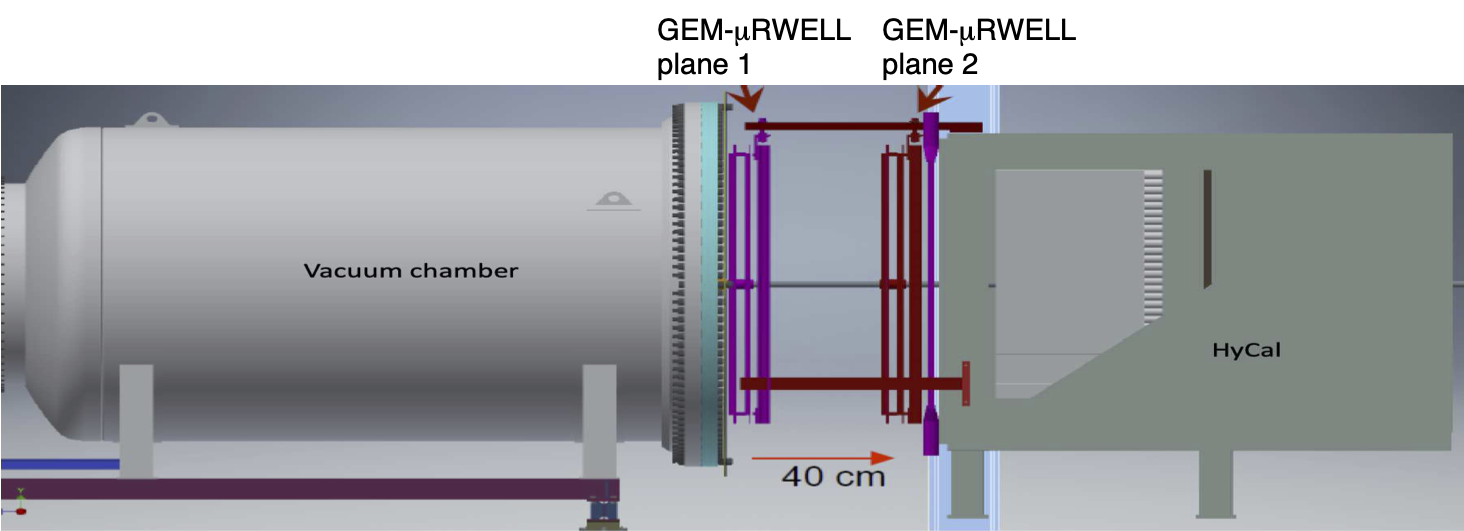}}
\caption{The placement of the new GEM-$\mu$Rwell chamber in the proposed experimental setup for PRad-II.}
\label{fig:PRad2_GEM}
\end{figure}

Just as in the PRad experiment the scattered electrons will travel through the 5 m long vacuum chamber with a  thin window to minimize multiple scattering and backgrounds. The vacuum chamber matches the geometrical acceptance of the calorimeter. The new  second GEM detector layer  will be placed about 40~cm upstream of the GEM detector layer location  in PRad, as shown in Fig.~\ref{fig:PRad2_GEM}. Both GEM layers will be made of  spacer-less detectors based on the novel$\mu$RWELL technology.  The pair of GEM-$\mu$RWELL detector planes will ensure a high precision measurement of the GEM detector efficiency needed for applying the integrated M{\o}ller method to the full angular range of the experiment. The two GEMs-$\mu$RWELL layers will also add a modest tracking capability to help further reduce the beam-line background. 

The elements of the experimental apparatus along the beamline are as follows:
\begin{itemize}
\item
windowless hydrogen gas target
\item 11$\times$ 11 cm$^2$ scintillator detector with a 4$\times$4 cm$^2$ hole placed 25 cm from the center of the target. 
\item
Two stage, large area vacuum chamber with a single thin Al. window at the calorimeter end
\item A pair of GEM-$\mu$RWELL detector planes, separated by about 40~cm for coordinate measurement as well as tracking.
\item
high resolution all \pbo ~crystal calorimeter (the Pb-glass part of the HyCal will be replaced with \pbo ~crystals) with fADC based readout.
\end{itemize}
\begin{figure}[!hbt]
\centerline{
\includegraphics[width=0.9\textwidth]{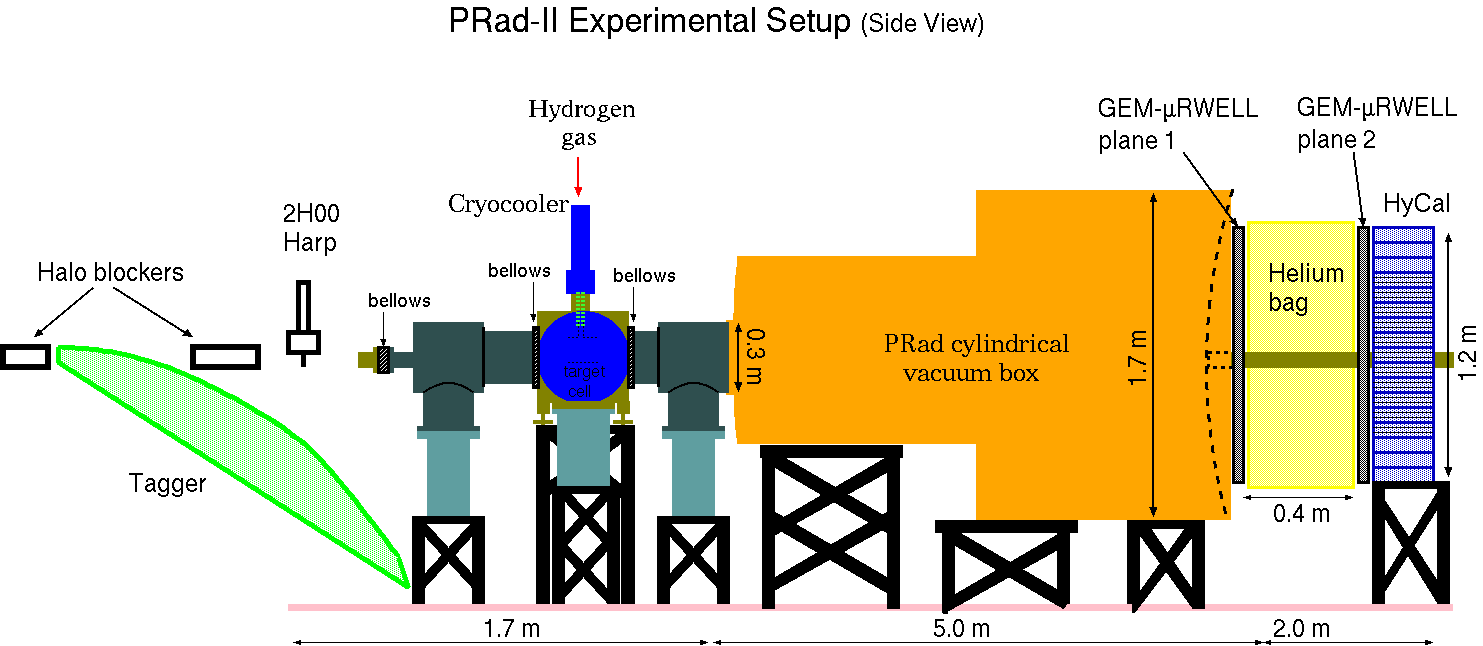}}
\caption{The proposed experimental setup for PRad-II.}
\label{fig:PRad2_setup}
\end{figure}
Figure~\ref{fig:PRad2_setup} shows a schematic layout of the PRad-II experimental setup.


\subsection{Electron beam}
\begin{figure}[htb!]
  \centerline{
    \includegraphics[width=0.53\textwidth]{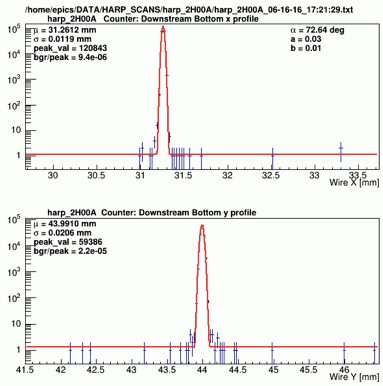}}
  \caption{Typical beam profile during the PRad experiment, showing a beam size of $\sigma_x=$~0.01~mm and $\sigma_y=$~0.02~mm.}
  \label{fig:beamprof}
\end{figure}

\begin{figure}[htb!]
  \centerline{
    \includegraphics[width=0.93\textwidth]{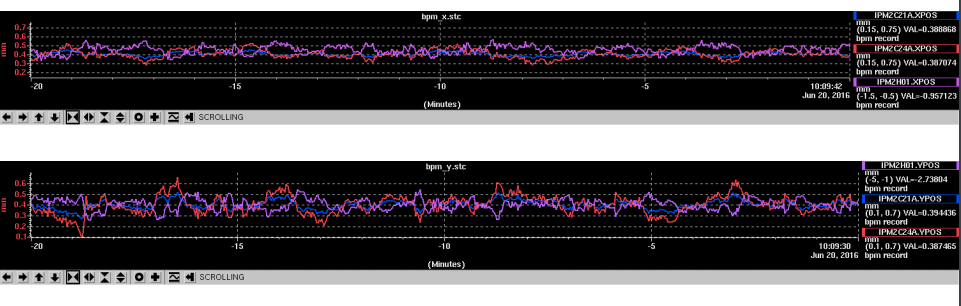}}
  \caption{Beam X,Y position stability ($\simeq \pm$ 0.1~mm) during the PRad experiment.}
  \label{fig:beampos}
\end{figure}

We propose to use the CEBAF beam at three incident beam energies $E_0 =$ 0.7, 1.4 and 2.1 GeV for this experiment. The beam requirements are listed in Table~\ref{tab:beampar}. All of these
requirements were achieved during the PRad experiment. A typical beam profile during the PRad experiment is shown in Fig.~\ref{fig:beamprof} and the beam X, Y position stability was  $\simeq \pm$ 0.1~mm as  shown in Fig.~\ref{fig:beampos}.

\begin{table}[hbt!]
\caption{Beam parameters for the proposed experiment}
\label{tab:beampar}
\vspace{-0.25cm}
\center{
\begin{tabular}{c|c|c|c|c|c} \hline \hline
  Energy & current & polarization & size & position stability & beam halo \\
  (GeV)  &   (nA)  &   (\%) &  (mm)  &  (mm) &  \\ \hline
  &         &        &        &       &  \\ [-11pt]
  0.7 & 20 & Non & $<$ 0.1 & $\leq$ 0.1 & $\sim$ 10$^{-7}$ \\
  1.4  &   70 & Non  & $<$ 0.1 & $\leq$ 0.1 & $\sim$ 10$^{-7}$ \\
  2.1  &   70 & Non  & $<$ 0.1 & $\leq$ 0.1 & $\sim$ 10$^{-7}$ \\ \hline \hline  
\end{tabular}
}
\end{table}


\subsection{Windowless hydrogen target}
\label{sec:target}
A critical component of this proposed experiment is a windowless hydrogen gas target used during the PRad experiment. PRad-II will reuse the windowless hydrogen gas flow target. This target had a thickness of \mbox{$\sim 2.5 \times 10^{18}$} hydrogen
atoms/cm$^2$, hence, with an incident beam current of 20 nA the luminosity is ${\cal L} \approx 3 \times 10^{29}$ cm$^{-2}$ s$^{-1}$.The high density was reached by flowing cryo-cooled hydrogen gas (at 19.5$^{\circ}$ K) through the
target cell with a 40~mm long and 63~mm diameter cylindrical  thin copper pipe. The upstream and downstream windows
of this cell were covered by  thin (7.5 $\mu$m) kapton films with 2~mm holes in the middle
for the passage of the electron beam through the target. Four high capacity turbo-pumps
were  used to keep the pressure in the chamber (outside the cell) at the
\mbox{$\sim 2.3$}~ mtorr level while the pressure inside the cell was
\mbox{$\sim 470$}~ mtorr. 
\begin{figure}[!ht]
\centerline{
\includegraphics[width=0.48\textwidth]{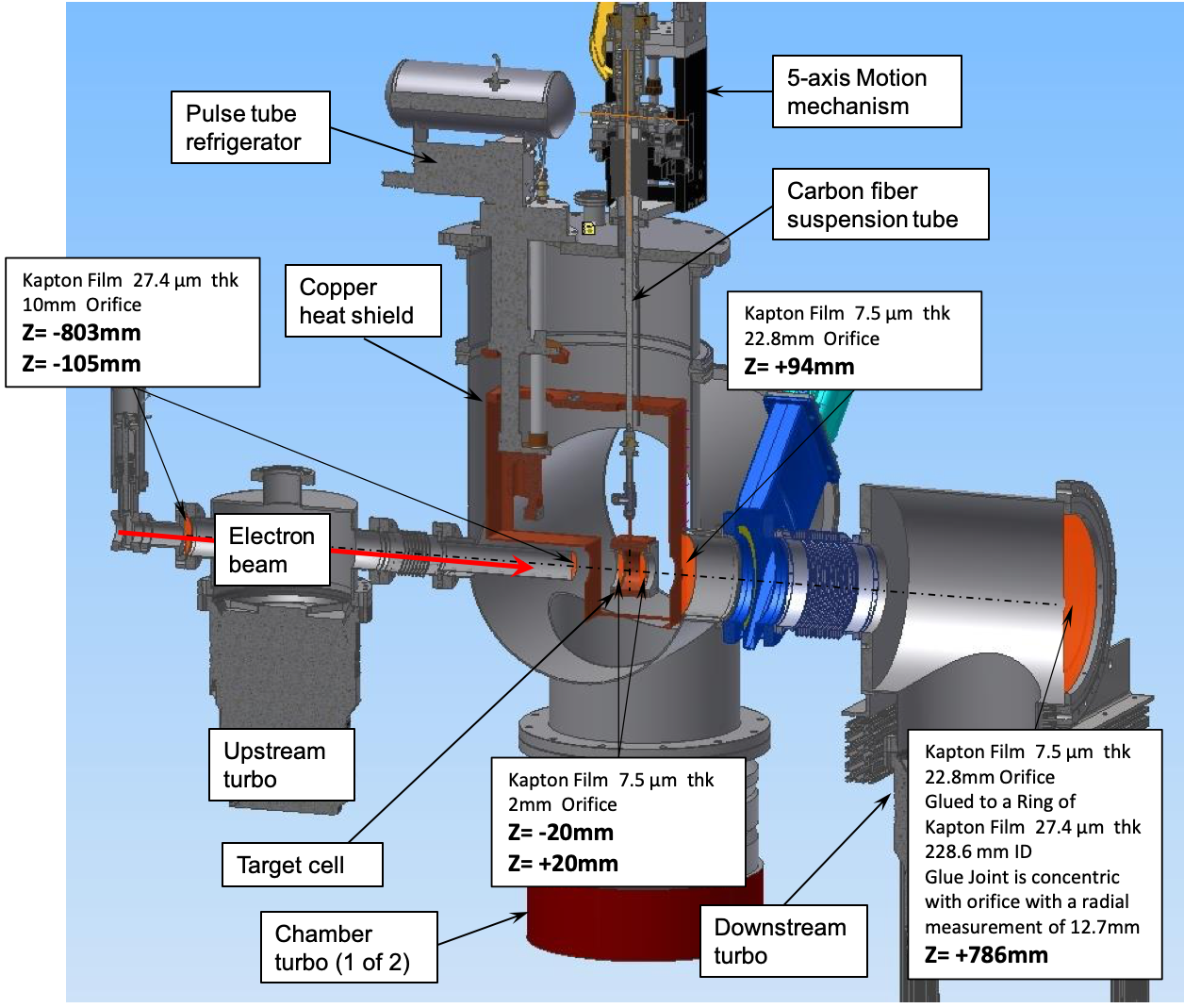}\hspace{0.04\textwidth}\includegraphics[width=0.48\textwidth]{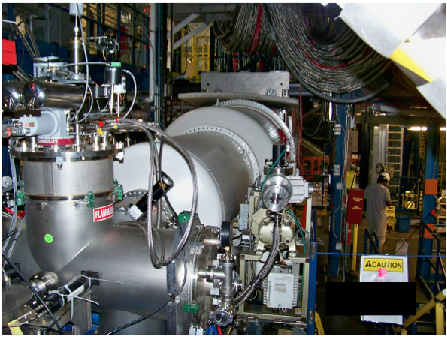}}
\caption{(left)Annotated drawing of the PRad gas flow target indicating most of the target’s main components. The location
and dimensions of various polyimide pumping orifices are shown, where Z is the distance from target center.
The direction of the electron beam is indicated by a red arrow. (right) Downstream view of the PRad target in the beamline.}
\label{fig:prad_target1}
\end{figure}

The target cell was specifically designed to create a
large pressure difference between the gas inside the cell and the surrounding beam line vacuum.

Figure~\ref{fig:prad_target1} (left) is a cut-thru drawing of the PRad target chamber and shows most of its major components. High-purity hydrogen gas ($>$99.99\%) was metered into the target system via a 0–10 slpm mass flow controller. Using a pair of remotely actuated valves, the gas was either directed into the target
cell for production data-taking, or into the target chamber for background measurements.
Before entering the cell, the gas was cooled to cryogenic temperatures using a two-stage pulse
tube cryocooler\footnote{Cryomech model PT810}  with a base temperature of 8 K and a cooling power of 20 W at 14 K. The
cryocooler’s first stage serves two purposes. It cools a tubular, copper heat exchanger that lowers
the hydrogen gas to a temperature of approximately 60 K, and it also cools a copper heat shield
surrounding the lower temperature components of the target, including the target cell itself. The
second stage cools the gas to its final operating temperature and also cools the target cell via
a 40 cm long, flexible copper strap. The temperature of the second stage was measured by a
calibrated cernox thermometer\footnote{Lakeshore Cryotronics}
 and stabilized at approximately 20 K using a small cartridge
heater and automated temperature controller. 


\begin{figure}[hbt!]
\centerline{
\includegraphics[width=0.3\textwidth]{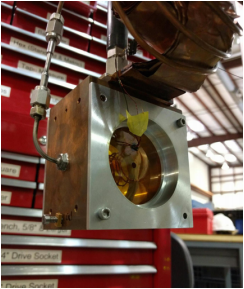}}
\caption{The PRad target cell. Hydrogen gas, cooled by the pulse tube cryocooler, enters the cell via the tube on the
left. The cell is cooled by a copper strap attached at the top, and is suspended by the carbon tube directly above the cell.
The 2 mm orifice is visible at the center of the polyimide window, as are the wires for a thermometer inside the cell. Two
1 $\mu$m solid foils of aluminum and carbon attach to the cell bottom, but are not shown in the photograph.}
\label{fig:target_cell}
\end{figure}

The target cell, shown in Fig.~\ref{fig:target_cell}, was machined from a single block of C101 copper. Its outer
dimensions are 7.5$\times$ 7.5 $\times$ 4.0 cm$^3$, with a 6.3 cm diameter hole along the axis of the beam line.
The hole is covered at both ends by 7.5 $\mu$m thick polyimide foils, held in place by aluminum
end caps. Cold hydrogen gas flows into the cell at its midpoint and exits via 2 mm holes at the
center of either kapton foil. The holes also allow the electron beam to pass through the H$_2$ gas
without interacting with the foils themselves, effectively making this a “windowless” gas target.
Compared to a long thin tube, the design of a relatively large target cell with small orifices on
both ends has two important advantages. First, it produce a more uniform density profile along
the beam path, allowing a better estimate of the gas density based upon its temperature and
pressure. Second, it eliminates the possibility of electrons associated with beam halo scattering
from the 4 cm long cell walls. Instead, the halo scatters from the 7.5 $\mu$m thick polyimide foils.
A second calibrated cernox thermometer, suspended inside the cell, provides a direct measure
of the gas temperature.  The
gas pressure was measured by a capacitance manometer located outside the vacuum chamber and
connected to the cell by a carbon fiber tube approximately one meter long and 2.5 cm in diameter.
The same tube is used to suspend the target cell, in the center of the vacuum chamber, from a
motorized 5-axis motion controller. The controller can be used to position the target in the path
of the electron beam with a precision of about $\pm$10 $\mu$m. It was also used to lift the cell out of the
beam for background measurements. Also,
two 1 $\mu$m thick foils, carbon and aluminum, were attached to the bottom of the copper target cell
for additional background and calibration measurements.
High-speed turbomolecular pumps were used to evacuate the hydrogen gas as it left the target
cell and maintain the surrounding vacuum chamber and beam line at very low pressure. Two pumps, each with a nominal pumping speed of 3000 l/s, were attached directly under the chamber, while pumps with 1400 l/s speed were used on the upstream and downstream portions of the beam line. A second capacitance manometer measured the hydrogen gas pressure inside the target chamber, while cold cathode vacuum gauges were utilized in all other locations. 

Polyimide pumping orifices were installed in various locations to limit the extent of high pres-
sure gas along the path of the beam. With this design, the density of gas decreases significantly outside the target cell, with 99\% of scattering occurring within the 4 cm length of the cell. 

\subsubsection{Target performance}
During the PRad experimet 600 sccm cold H$_2$ gas was flown through the target cell. Under these conditions,
typical pressure and temperature measurements inside the target cell were 0.48 torr and 19.5 K,
respectively, resulting in a gas density of 0.83 mg/cm$^3$~\cite{nistgas}. Table~\ref{tab:gas_pressure} gives typical pressure
measurements obtained in other regions of the electron beam path. The hydrogen areal density
is calculated as the product of the gas number density and the length of the region. In all regions
except the target cell, a room temperature of 293 K is assumed when calculating the gas density.
The vast majority of the hydrogen  gas  was confined to the 4 cm long target cell, with the majority of the
remaining gas being measured in the 5 m long, 1.8 m diameter vacuum chamber just upstream of
the calorimeter. Here the achievable vacuum pressure was limited by the conductance between
the chamber and its vacuum pump.
\begin{table}[!hbt]
  \caption{Hydrogen gas pressures and areal densities for the PRad beam line. Refer to Fig.~\ref{fig:prad_target1} (left) for more details.Room temperature gas is assumed in calculating the areal density of all regions except Region 1 (target cell), where a temperature of 19.5 K was used.}
  \label{tab:gas_pressure}
\center{
\begin{tabular}{|c|c|c|c|c|}\hline
Region & Length & Pressure & Areal density & Percentage of total \\ 
& (cm) & (torr) & (atoms/cm$^2$) &  \\\hline
Target cell & 4 &0.48 &1.9 $\times$ 10$^{18}$& 98.97 \\
Upstream beamline  & 300 & 2.2 $\times$ 10$^{-5}$& 2.0 $\times$ 10$^{14}$& 0.02 \\
Upstream chamber & 71& 5.7 $\times$ 10$^{-5}$ &2.6 $\times$ 10$^{13}$& 0.00 \\ 
Target chamber& 14 &2.3 $\times$ 10$^{-3}$& 2.1 $\times$ 10$^{15}$& 0.11 \\
Downstream chamber  & 71& 3.0 $\times$ 10$^{-4}$ &6.1 $\times$ 10$^{14}$& 0.07 \\
Vacuum chamber & 400& 5.2 $\times$ 10$^{-4}$ &7.2 $\times$ 10$^{15}$ &0.83 \\ \hline
\end{tabular}
}
\end{table}
Two types of background measurements were made. In the first, the H$_2$ gas flow was maintained at the same 600 sccm, but the gas was directed into the vacuum chamber rather than the target cell. In this case, the chamber pressure increased slightly to 2.9 mtorr, and the cell temperature warmed to 32 K.  For the second
type of background measurements, the gas flow was set to zero, in which case both the cell and
chamber pressures dropped below 0.001 torr.
\begin{figure}[!ht]
\centerline{
\includegraphics[width=0.65\textwidth]{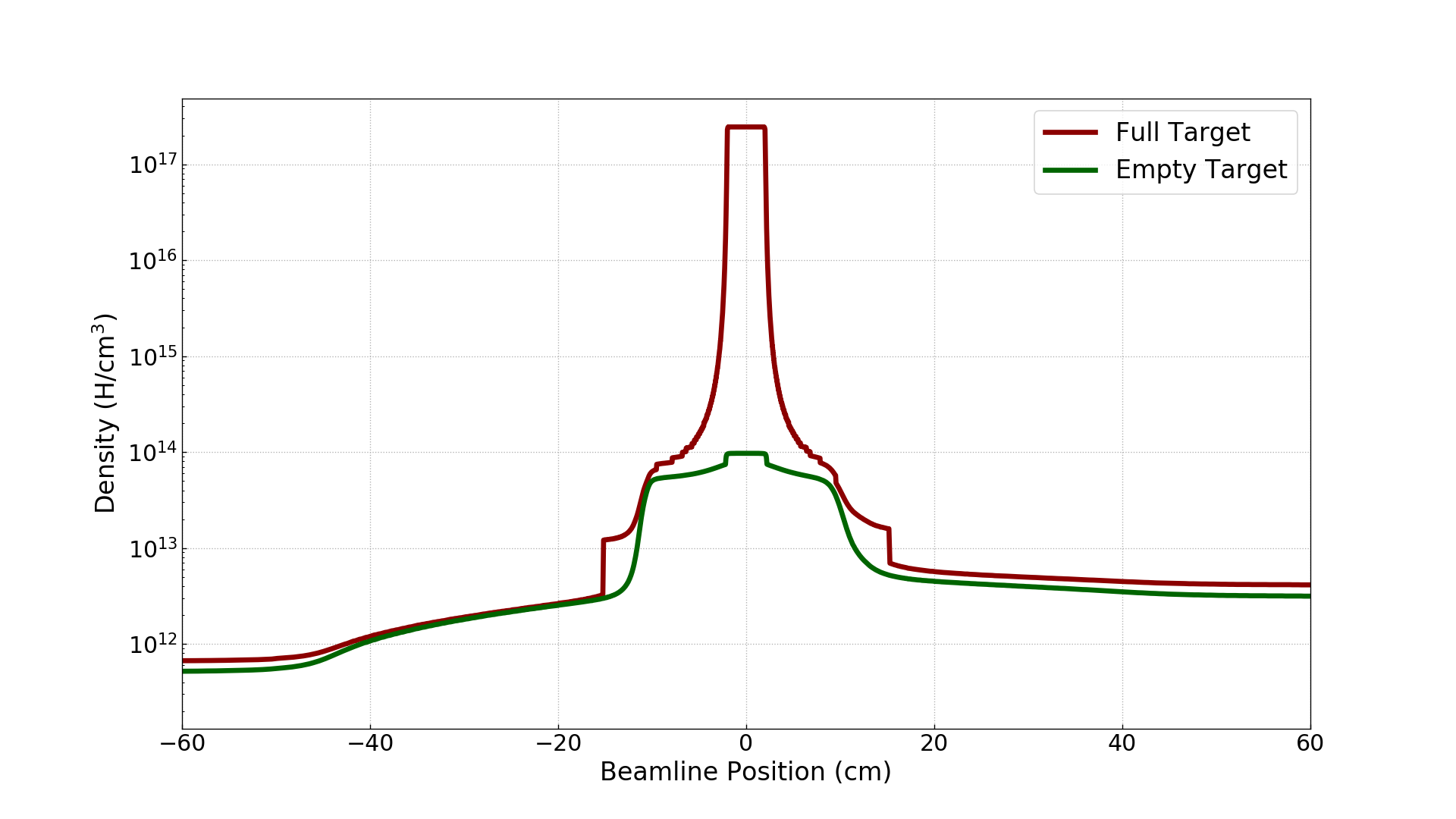}}
\caption{Density profile of hydrogen atoms along the electron beam line. Here, the target cell is centered at 0 cm, and
the electron beam transverses the target from negative to positive values. The red line indicates a measurement with
600 sccm of hydrogen flowing into the target cell. The green line indicates a background measurement with the same
flow of gas directly into the target vacuum chamber.}
\label{fig:comsol_sim}
\end{figure}

\begin{figure}[!ht]
  \centerline{
    \includegraphics[width=0.5\textwidth]{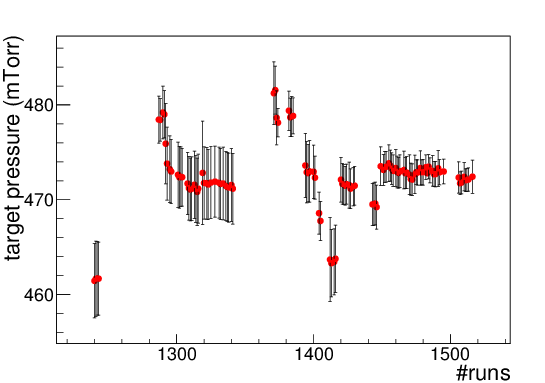}\includegraphics[width=0.5\textwidth]{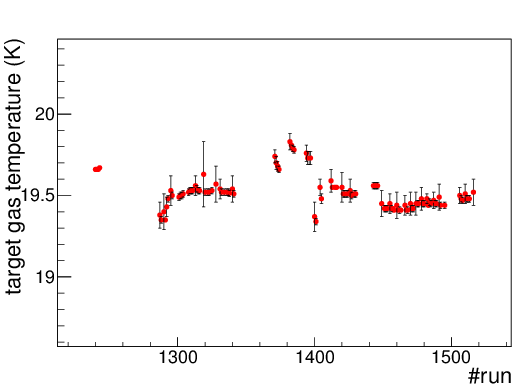}}
  \caption{The variation of PRad target pressure and temeperature {\it vs.} run number. Each run was about 1~hr long.}
  \label{fig:stable}
\end{figure}

The measured temperature values, together with the inlet gas flow rate, pumping speeds of the pumps, and the detailed geometry of the target system were used to simulate the hydrogen density profile in the target using the COMSOL Multiphysics \textsuperscript{\textregistered} simulation package. The average pressure obtained from the simulation agreed with the measured values within 2~mTorr for both the target cell and the target chamber, under the PRad production running conditions. Fig.~\ref{fig:comsol_sim} shows the simulated density profile along the beam path for both the full target cell configuration
and the “full chamber” background configuration.  
During the PRad experiment the target pressure and temperature remained stable throughout. The variation of target pressure and
temperature with time is shown in Fig.~\ref{fig:stable}. 

\subsection{Large volume vacuum chamber}
\begin{figure}[!ht]
  \centerline{
    \includegraphics[width=0.65\textwidth]{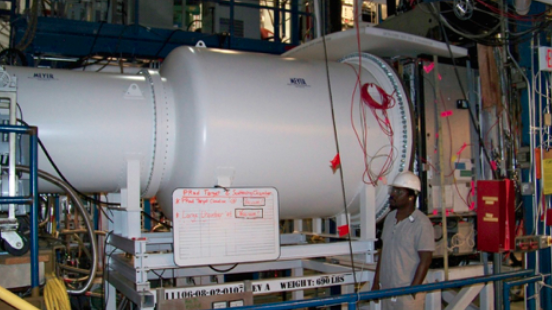}\hspace{5ex}\includegraphics[width=0.25\textwidth]{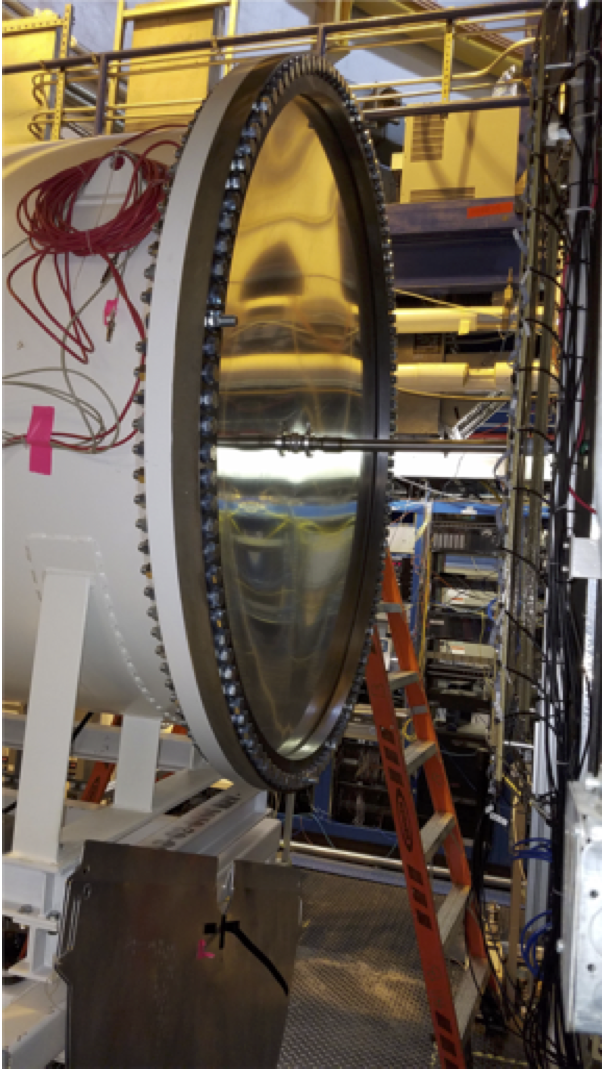}}
  \caption{A photograph of the $\sim$5~m long, two stage vacuum chamber used during the PRad experiment (left). A photograph of the 1.7~m diameter thin window
  at one end of the vacuum chamber (right). Here the GEM and HyCal have been moved downstream for technical service.}
  \label{fig:vacphoto}
\end{figure}

For the PRad experiment a new large $\sim$5~m long, two stage vacuum chamber was designed and built. It extended from the target to the GEM/HyCal detector system.
There was a single $1.7$~m diameter, $63$ mil thick Al. window at one end of the vacuum chamber, just before the GEM detector. A 2-inch diameter beam pipe was
attached using a compression fitting to the center of the thin window. This design ensured that the electron beam did not encounter any additional material
other than the hydrogen gas in the target cell, all the way down to the Hall-B beam dump. The vacuum box also helped minimize multiple scattering of the scattered electrons en route to the detectors. A photograph of the vacuum chamber is shown in Fig.~\ref{fig:vacphoto}.
This vacuum chamber will be reused for PRad-II.

\subsection{High resolution forward calorimeter}
\label{sec:calor}

The scattered electrons from $e-p$ elastic and M{\o}ller scatterings
in this precision experiment will be detected with a high resolution 
and high efficiency electromagnetic calorimeter.
In the past decade, lead tungstate (\pbo) has became a popular inorganic 
scintillator material for precision compact electromagnetic calorimetry 
in high and medium energy physics experiments (CMS, ALICE at the LHC) 
because of its fast decay time, high density and high radiation hardness. 
The performance characteristics of the \pbo ~crystals are well known mostly 
for high energies ($>$10 GeV)~\cite{CMS94} and at energies below one 
GeV~\cite{Mengel98}.
The PrimEx Collaboration at Jefferson Lab constructed a
novel state-of-the-art multi-channel electromagnetic hybrid (\pbo-lead glass)
calorimeter (HYCAL)~\cite{primex-cdr} to perform a high precision (1.5\%) 
measurement of the neutral pion lifetime via the Primakoff effect.
The advantages of using the HyCal calorimeer was also demonstrated in the PRad experiment.

\begin{figure}[!ht]
\centerline{
\includegraphics[width=0.48\textwidth]{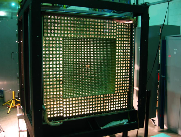}}
\caption{The PrimEx HYCAL calorimeter with all modules of the high
performance \pbo ~crystals in place.}
\label{fig:hycal}
\end{figure}

For PRad-II we are proposing to replace the outer Pb-glass layer with \pbo ~modules turning the calorimeter into a fully \pbo ~calorimeter. A single \pbo ~module is $2.05 \times 2.05$ cm$^2$ in cross sectional
area and 18.0 cm in length (20$X_0$).
The calorimeter consists of 1152 modules arranged in a $34 \times 34$
square matrix ($70 \times 70$ cm$^2$ in size) with four crystal detectors 
removed from the central part ($4.1 \times 4.1$ cm$^2$ in size) for passage 
of the incident electron beam. An additional $\sim$ 1500 modules will be used to replace the $\sim$ 800 Pb-glass modules.
As the light yield of the crystal is highly temperature dependent
($\sim 2$\%/$^{\circ}$C at room temperature), a special frame was developed
and constructed to maintain constant temperature inside of the calorimeter
with a high temperature stability ($\pm 0.1^{\circ}$C) during the experiments.
Figure~\ref{fig:hycal} shows the assembled PrimEx HYCAL calorimeter that was used in the PRad experiment.  
For the PRad-II experiment the calorimeter will be placed at a distance of about 5.5 m from the target which will provide a geometrical acceptance of about 25 msr.

\begin{figure}[!ht]
  \centerline{
    \includegraphics[width=0.48\textwidth]{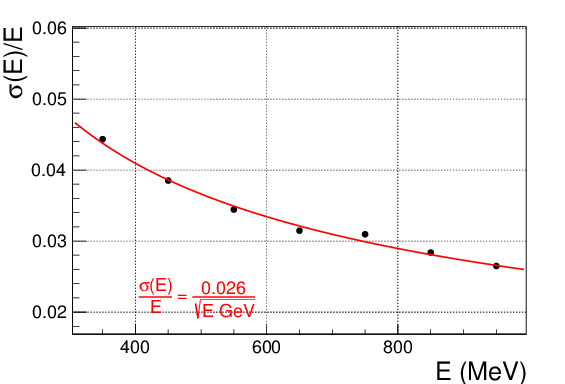}\includegraphics[width=0.48\textwidth]{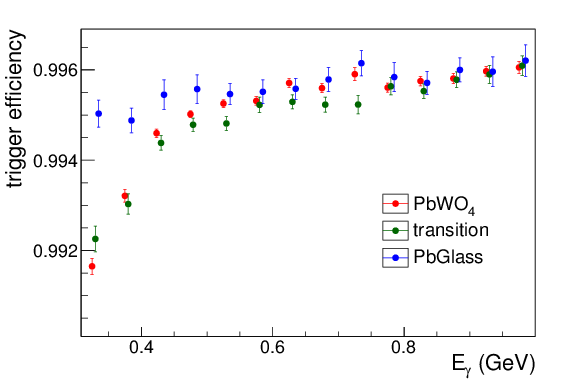}}
  \caption{Energy resolution of the PbWO$_4$ crystal part of the HyCal calorimeter (left) and the energy dependence of the trigger efficiency (right). These data are from the PRad experiment.}
  \label{fig:hycal_calib}
\end{figure}

During PRad the energy calibration of HyCal was performed by continuously irradiating the
calorimeter with the Hall B tagged photon beam at low intensity ($<$ 100 pA).
 An excellent energy resolution of $\displaystyle \sigma_E/E = 2.6\%/\sqrt{E}$
has been achieved by using a Gaussian fit of the line-shape obtained from the
$6 \times 6$ array.
The impact coordinates of the electrons and photons incident on the crystal
array were determined from the energy deposition of the electromagnetic shower
in several neighboring counters.
Taking into account the photon beam spot size at the calorimeter
($\sigma$=3.0 mm), the overall position resolution reached was
$\sigma_{x,y} = 2.5 ~{\rm mm}/\sqrt{E}$ for the crystal part of the
calorimeter. The calorimeter performed as designed during the experiment, as shown in Fig.~\ref{fig:hycal_calib}, which shows
the resolution achieved during the PRad experiment and the energy dependence of the trigger efficiency.

\begin{figure}[hbt!]
  \centerline{
    \includegraphics[width=0.65\textwidth]{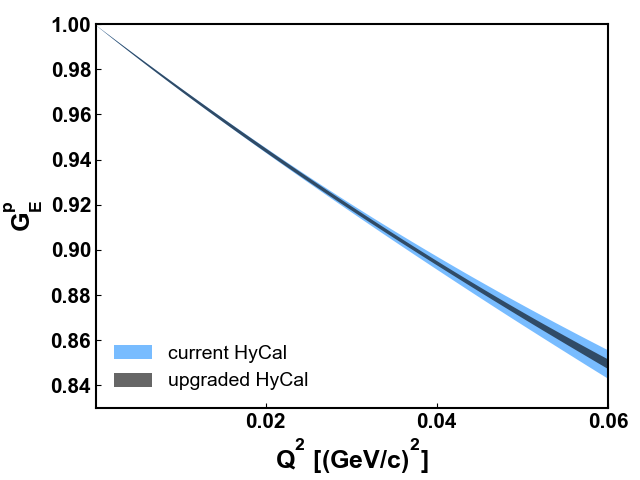}}
  \caption{The one standard deviation systematic uncertainty band in the extracted $G_{E}^{p}$ for the current HyCal and the upgraded calorimeter.}
  \label{fig:hycal_upgrade}
\end{figure}

The upgraded calorimeter will provide enhanced uniformity across the entire calorimeter and reduce the uncertainty due to $e-p$ inelastic contribution to the elastic $e-p$ yield (event selection). 
The impact of the upgraded HyCal on the uncertainty in event selection and detector response was studied using the PRad comprehensive  Monte Carlo simulation. Fig.~\ref{fig:hycal_upgrade} shows the projected improvement in the one standard deviation systematic uncertainty band in the extracted $G_{E}^{p}$.   
The trigger for the PRad-II experiment will be total energy deposited in the calorimeter $\ge$ 20\% of $E_0$. This will allow for the detection of the \moll ~events in both single-arm and double-arm modes.

\subsection{GEM $\mu$RWELL based coordinate detectors}
\label{sec:GEM}
\begin{figure}[!hbt]
\centerline{
\includegraphics[width=0.35\textwidth]{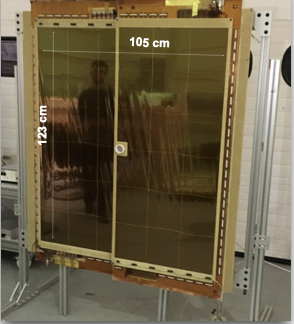} \includegraphics[width=0.45\textwidth]{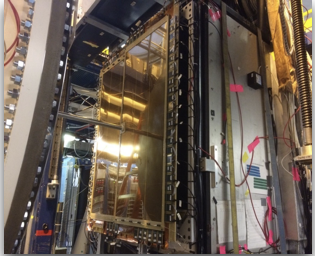}}
\caption{The PRad GEM chambers (left) and the GEM chambers mounted on the HyCal during the experiment (right).}
\label{fig:prad-gems}
\end{figure}
The PRad experiment used Gas Electron Multiplier (GEM) based coordinate detectors with $\sim$~56~$\mu$m position resolution. 
The active area of the GEM PRAd layer was 123 cm $\times$ 110 cm to match the area of the calorimeter. The GEM layer was made of two large area GEM detectors,  each with an  active area of 123 cm $\times$ 55 cm, arranged so that there is a narrow overlap area in the middle. An especially designed through hole with a 4 cm radius  built into  GEM detectors at the center of the  active area allowed for the passage of the beamline.  The  GEM detectors were  triple GEM foil structures followed by a 2D x-y strip readout layers. The chambers were mounted to the front face of the HyCal calorimeter using a custom mounting frame.
A pre-mixed gas of 70\% Argon and 30\% CO$_2$ was continuously supplied to the chambers. Figure~\ref{fig:prad-gems} shows the PRad GEM detector and a view of it mounted to the front of the HyCal calorimeter during the PRad experiment.

The chambers were designed and constructed
by the University of Virginia group and are currently the largest such chambers to be used in a nuclear physics experiment. These GEM
chambers provided more than a factor of 20 improvement in coordinate resolution and a similar improvement in the $Q^2$ resolution. They allowed unbiased coordinate reconstruction of hits on the calorimeter, including the transition region of the HyCal calorimeter. The GEM detectors also allowed us to use the lower resolution
Pb-glass part of the calorimeter, extending the total $Q^2$ range covered at a single beam energy setting.

\begin{figure}[!ht]
  \centerline{
    \includegraphics[width=0.5\textwidth]{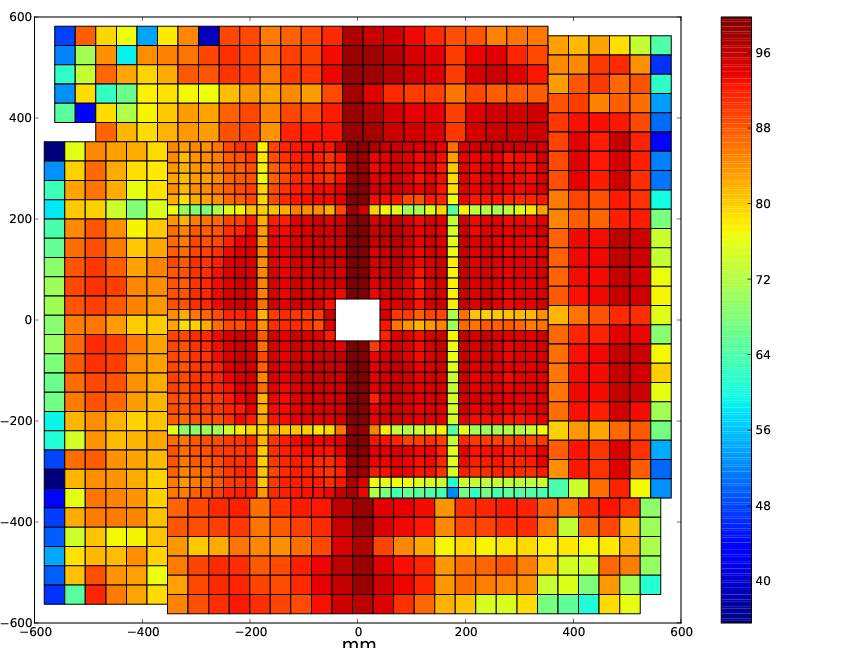}\includegraphics[width=0.5\textwidth]{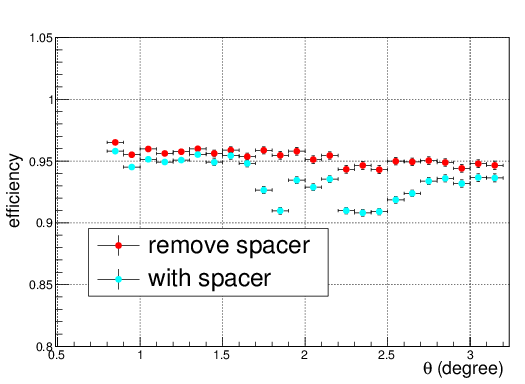}}
  \caption{A plot of the GEM efficiency over the X-Y coordinates of the detector (left), and the GEM efficiency over the region overlapping with the~\pbo~crystals of the HyCal
    calorimeter {\it vs.} polar angle (right). The drops in efficiency seen in the 2D plot in the left is due to spacers
    inside the GEM modules.  A software cut to remove the spacers yields an efficiency profile uniform to within +/- 1\% level as seen by red circles.
    The cut to remove spacers reduce the available statistics by only about 4.7\%.}
  \label{fig:gem_effres}
\end{figure}

\begin{figure}
\centerline{
\includegraphics[width=0.52\textwidth]{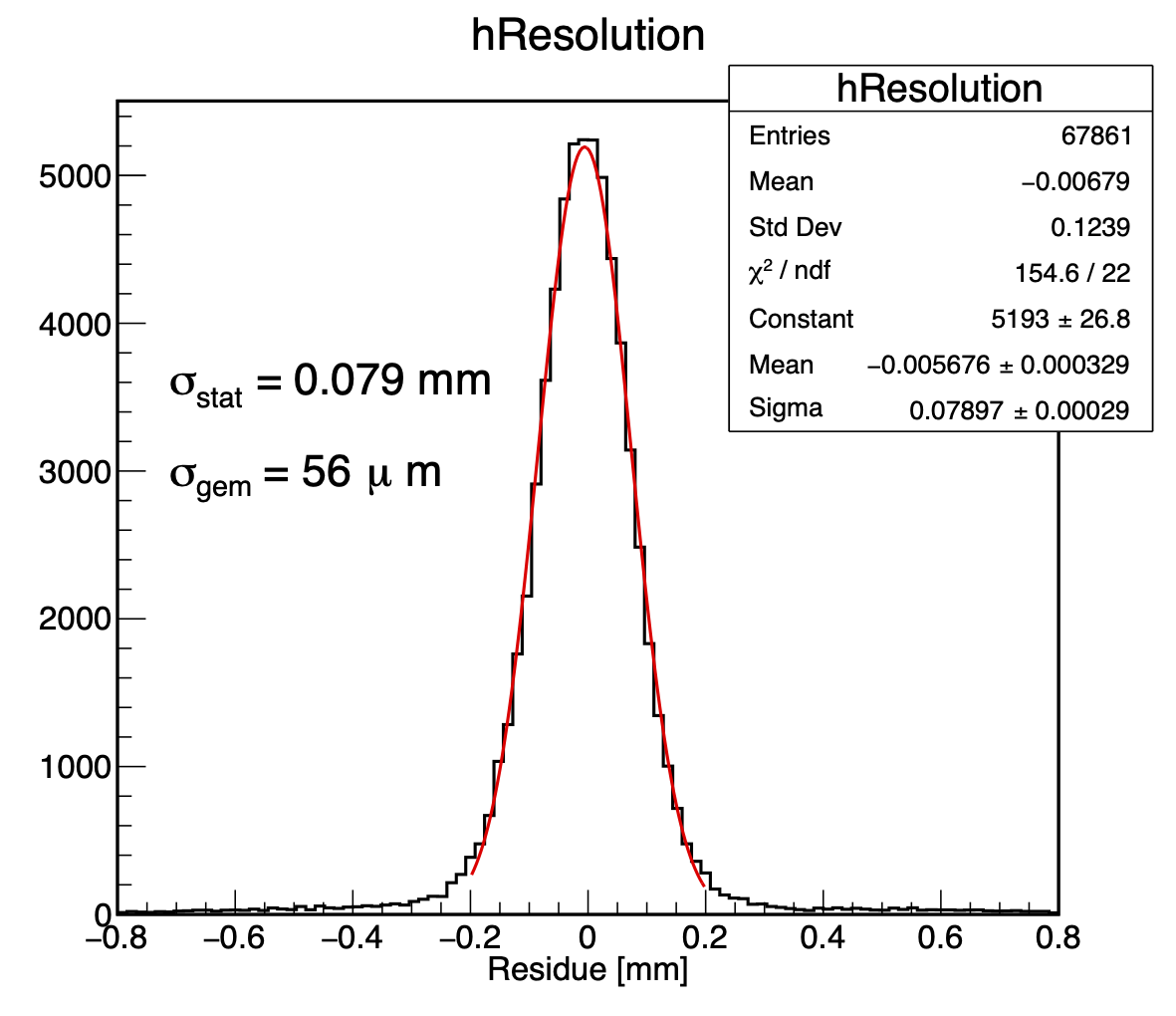}
\includegraphics[width=0.48\textwidth]{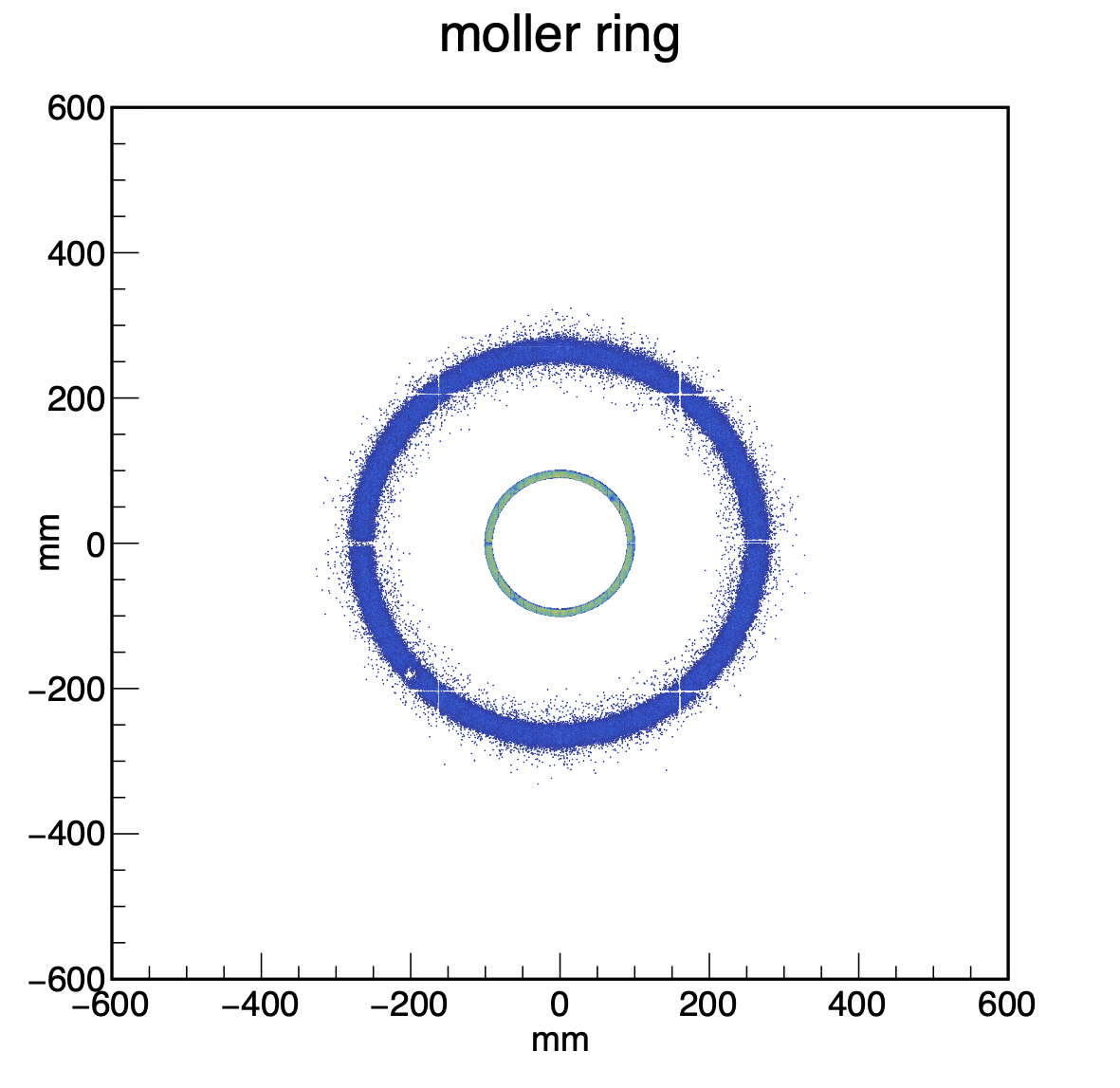}}
\caption{(Left)  The  position resolution (approximately 56 $\mu$m) for GEM detectors achieved during PRad experiment; this represents a factor of 20-40 improvement over the resolution available without the GEM tracker in the setup. (Right) The scattered M{\o}ller {\it ee} pair rings detected by PRad GEM tracker illustrating the high position resolution and accuracy provided by the GEMs. Furthermore, this plot shows the very low background level in the reconstructed GEM hit locations.  }
\label{prad-data}
\end{figure}

The PRad GEM detectors were readout using the APV25 chip
based Scalable Readout System (SRS) developed at CERN.   An upgraded firmware  configuration developed for the  PRad setup allowed the experiment to collect data
at $\sim$ 5kHz with a data rate of $\sim$ 400~MB/sec and $\sim$ 90\% live time. This was the highest DAQ rate achieved by a APV based system at the time. 

The PRad GEM  detectors consistently performed well throughout the experiment. The efficiency of the chamber
was mostly uniform over the entire chamber,  except for over the spacer locations,  as shown in Fig.~\ref{fig:gem_effres}, and it achieved the design resolution of $<$ 70 $\mu$m.  The performance of the detector remained stable throughout the experiment. In the PRad GEM chambers the  2$\times$4 grid of thin dielectric spacers was used between each pair of GEM foils to prevent them from coming into contact with each other. Each large area GEM foil was sub-divided into 60 sectors; the 61.5 cm long and 18.3 cm wide sectors were separated by narrow (100 $\mu$m) margins. The GEM  efficiency loss due to the presence of spacers and sector margins was measured relative to HyCal  using data and was modeled in the simulation.

\begin{figure}[hbt!]
\includegraphics[width=0.625\textwidth]{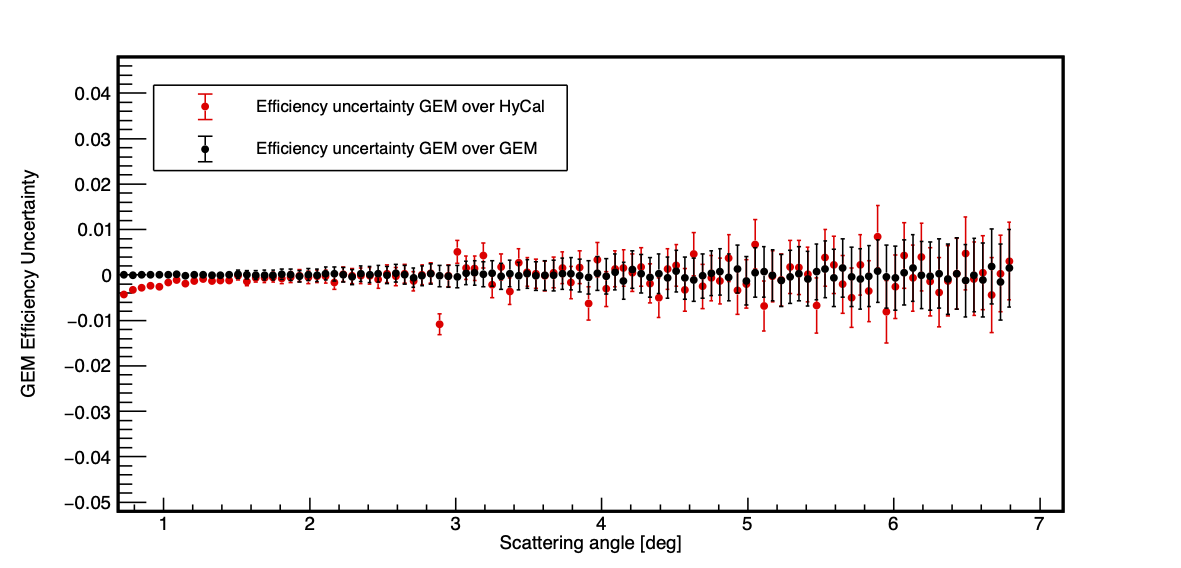}\includegraphics[width=0.375\textwidth]{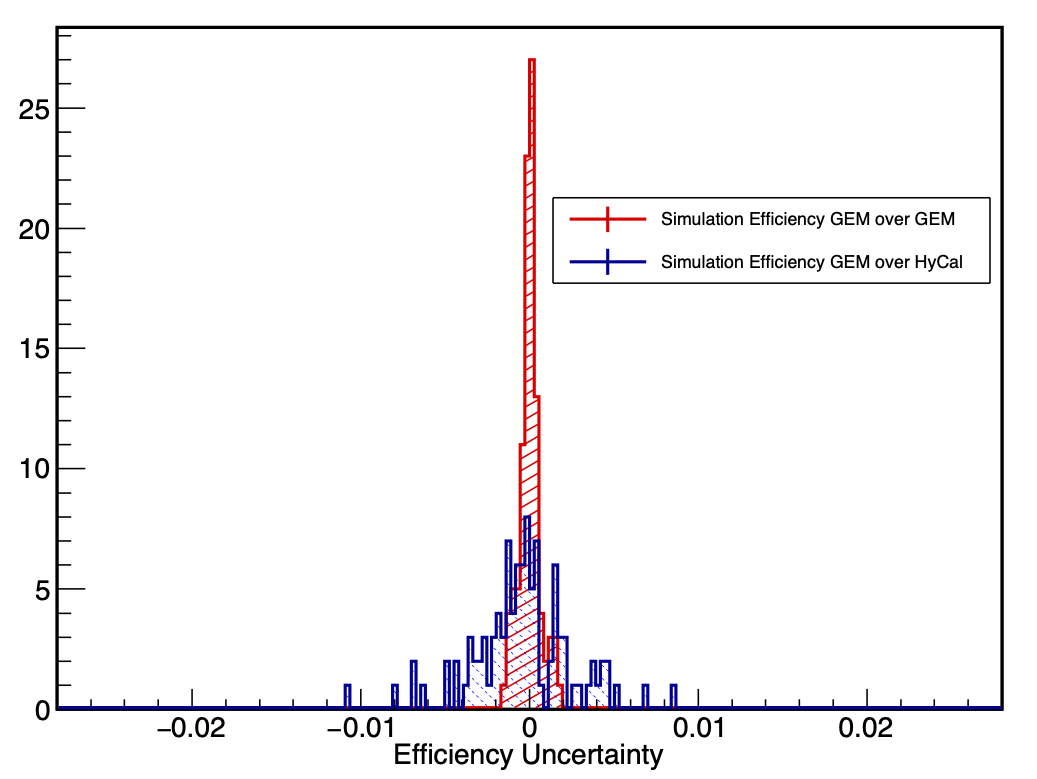}

\caption{(left) Simulated GEM efficiency uncertainty as a function of scattering angle, when using a single GEM detector plane along with the HyCal compared to when using two spacer-less GEM-$\mu$RWELL detector planes. (right) The uncertainty in determining the efficiency for single GEM$-\mu$RWELL vs two GEM-$\mu$RWELL detector planes.}
\label{fig:GEM_new_eff}
\end{figure}

\begin{figure}[hbt!]
\centerline{\includegraphics[width=0.5\textwidth]{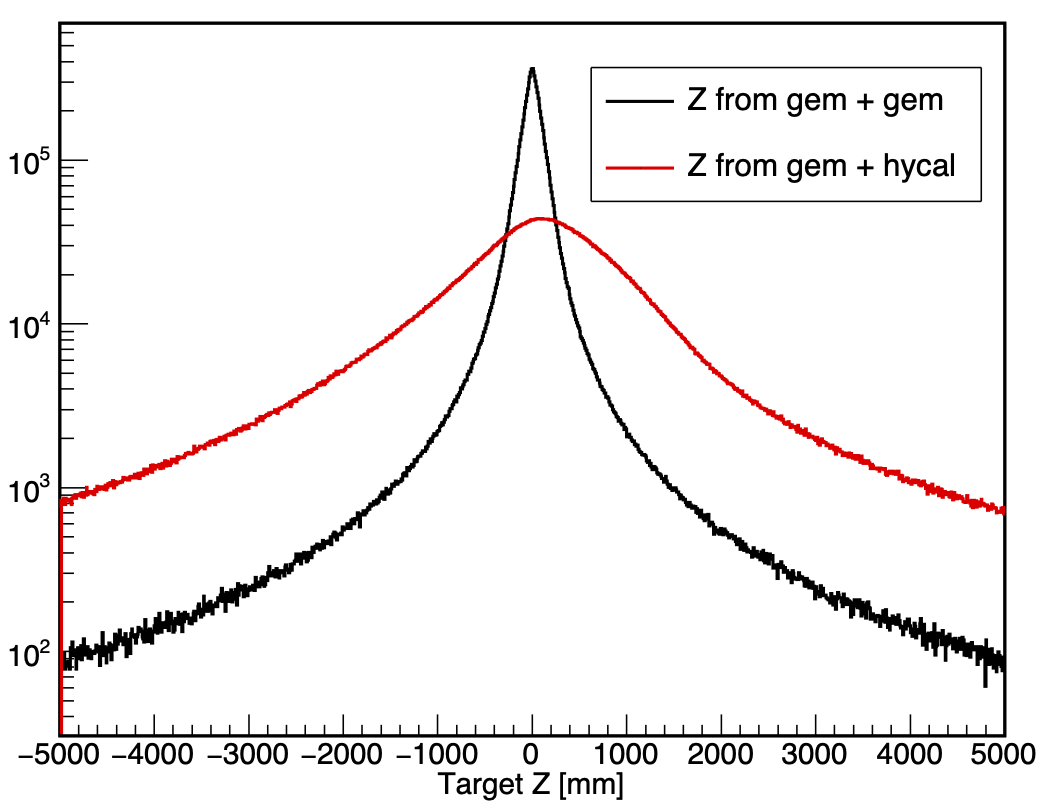}}
≈
\caption{Reconstructed reaction z-vertex when using one GEM plane along with the HyCal vs using two GEM-$\mu$RWELL detector planes.}
\label{fig:new_z_recon}
\end{figure}

The new  $\mu$RWELL  based tracking layers will have an identical size and outer design to the PRad GEM detectors. However, new advances in $\mu$RWELL  detector technology such as spacer-free construction with a smaller materials budget will be incorporated into the new  detectors. The
impact  of using two advanced technology coordinate detector layers  on the determination of   inefficiency  profile and the associated uncertainty,  as well as the improvement in the vertex reconstruction capabilities was studied using a simulation of the GEM detectors. The improvement in the determination of the efficiency and its uncertainty is shown in Fig.~\ref{fig:GEM_new_eff}. In addition the improvement in the resolution of the reconstructed reaction vertex is shown in Fig.~\ref{fig:new_z_recon}.

 The readout of the two  GEM $\mu$RWELL layers requires approximately 20 k electronic channels.  This readout for the proposed experiment will be done by using the high-bandwidth optical link based MPD readout system recently   developed  for the SBS program in Hall A.  This system is currently under rigorous resting. This new system uses the APV-25 chip used in the PRad GEM readout. However, the readout of the digitized data is performed over a high-bandwidth optical link to a  Sub-System Processor (SSP) unit in a CODA DAQ setup.  Given its 40 MHz  sampling rate and the number of multiplexing channels, the  limiting  trigger rate for the APV chip  is  280 kHz in theory.  In practice we expect it to be lower and assume a 100 kHz limit.  Currently tests are underway by the JLab electronics group in collaboration with the UVa group to test the SBS GEM readout system to operate at this high trigger rate limit. Given the aggressive R\&D program currently in place to reach this goal, we do not anticipate any difficulty of reaching the 25 kHz trigger rate assumed for the PRad-II experiment. 
 
 The option for an even faster GEM readout system is now available with the currently ongoing work as part of the pre R\&D program for Jefferson Lab Hall A SoLID project. This fast GEM readout system is based on the new  VMM chip was  developed at BNL for the ATLAS large Micromegas Muon Chamber Upgrade. VMM chip  is an excellent candidate for large area Micro Pattern Gaseous Detectors such as GEM and $\mu$RWELL detectors.
The VMM is a rad-hard chip with  64 channels with  an embedded ADC for each channel. This chip is especially suited for high rate applications  and is much more advanced than the 20  year old APV chip.
The VMM  chip has an adjustable shaping time which can be set to be as low as 25 ns.
In the standard (slower) readout mode, the ADC provides 10-bit resolution, while in the faster, direct readout mode the ADC resolution is limited to 6-bits.
The fast direct readout mode has a  very short circuit-reset time of less than 200 ns following processing of a signal. The VMM chip has already been adapted by the CERN RD-51 collaboration for Micro-Pattern Gas Detectors  to replace the APV-25 chip. The electronics working group of the RD-51 collaboration  has already created a new version of its Scalable Readout System (SRS) based on the VMM chip. The UVa  group, which  has extensive expertise  operating  the APV based SRS readout,  recently acquired  a 500 channel  VMM-SRS system and is testing it in collaboration with the Jlab DAQ group.  Furthermore,  the as part of the SoLID pre R\&D program the Jlab electronics group is now developing a GEM readout system capable of running at 300 kHz based on the VMM chip. 

The 170 k channel  APV based GEM readout for the HallA SBS project has been  already acquired and built,  while as part of the HallA SoLID project,  a 200+ k channel VMM based readout system will be assembled. Given these very large volume fast  readout systems, we do not see any problem acquiring the 20 k channel GEM readout system needed for PRad-II

\begin{figure}
\centerline{
\includegraphics[width=16cm, angle = 0]{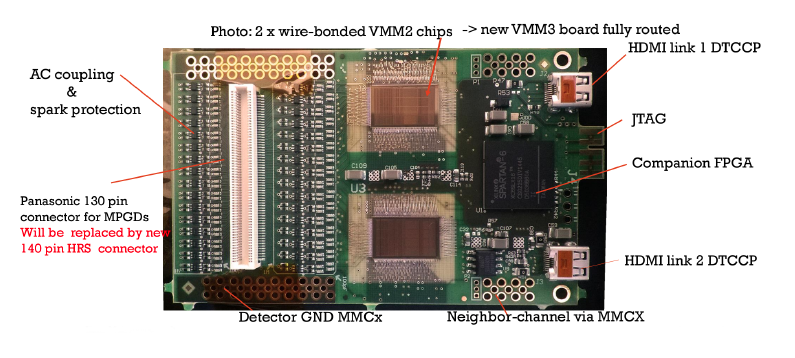}}
\caption{ The VMM chip based CERN RD-51 SRS readout card. The previous generations of this readout card (for example  the card used for PRad) were based on the APV-25 chip.}
\label{VMM_SRS}
\end{figure}

\subsection{Electronics, data acquisition, and trigger}
\label{subsec:trigger}

The high resolution calorimeter in this proposed experiment will  have around 2500 channels of charge and timing information.  These  will be readout using the JLab designed and built flash-ADC modules (FADC250),  each with  16 channels. The DAQ system for the calorimeter is thus composed of 160 FADC250 modules that can be held in ten 16-slots VXS crates. The major advantages of the flash-ADC based readout are the simultaneous pedestal measurement (or full waveform in the data stream), sub-nanosecond timing resolution, fast readout speed, and the pipeline mode that allows more sophisticated triggering algorithms such as cluster finding.

Additionally, some VME scalers will read out and periodically inserted into the data stream.

The DAQ system for the proposed experiment is the standard JLab CODA
based system utilizing the JLab designed Trigger Supervisor.
A big advantage of the CODA/Trigger Supervisor system is the ability
to run in fully buffered mode.
In this mode, events are buffered in the digitization modules themselves
allowing the modules to be ``{\it live}'' while being readout.
This significantly decreases the deadtime of the experiment. 
With the upgraded flash-ADC modules we expect to reach a data-taking rate of about 20 kHz 
events, which is about $4$ times higher than the data-taking rate 
in PRad experiment. 
Such a capability of the DAQ system has already been demonstrated by CLAS12 experiments.


A large fraction of the electronics needed for the PRad-II DAQ and 
trigger, including the high voltage crates and all necessary cabling for 
the detectors, are available in Hall B from the PRad experiment.

Our approach in organizing the first level hardware trigger in this
proposed experiment is to make it as simple as possible to reach
the highest efficiency for the event selection process and in the
mean time, to meet the DAQ rate requirements.
The primary trigger will be formed from the \pbo ~calorimeter by
only using the analog sum of all dynode outputs from each of the
crystal cells.

The scattered electrons from the \epreac ~reaction carry almost the
same energy as the incident beam.
Therefore, for this process alone, one can organize a very efficient
trigger by requiring the total energy in the calorimeter to be
$0.8 \times E_0$ including the resolutions.
We are planning to detect simultaneously the electrons from the
\mollreac ~process in this experiment in two single-arm and
coincidence modes.
For the coincidence mode, we are required to lower the total
energy threshold level to about one-fifth of the beam 
energy $-$ $0.2 \times E_0$ 
including the resolutions.
This will be still reasonable for this low luminosity 
$({\cal L} \approx  3 \times 10^{29}$ cm$^{-2}$ s$^{-1})$
and low background experiment.

\subsection{Improved Radiative corrections at forward angles}
\label{sec:newrad}

In order to reach a high precision in proton radius experiments such as PRad \cite{Xiong:2019,Xiong:2020}, in addition to a tight control of systematic
uncertainties and a precise knowledge of backgrounds associated with the experiment, a careful calculation of radiative corrections (RC) is necessary. It should be noted that the RC calculations carried out for small scattering angles give radiative corrections that are smaller than the corrections obtained from larger angles. Consequently, small angle scattering experiments like PRad/PRad-II, in this respect have a significant advantage as compared to other scattering experiments performed at larger angles.

Since in the PRad experiment both elastic $e-p$ and M{\o}ller $e-e$ scattering events are taken simultaneously during the experiment, the integrated luminosity is canceled out in the ratio between the two differential cross sections since it is the same for both reaction channels. However, one also needs to take into account that an experimental differential cross section cannot be used directly for a form factor extraction, as it contains radiative effects. To obtain the Born level differential cross section at a particular angle, one needs to apply precisely calculated RC to the cross section and also include a systematic uncertainty associated with the calculation.

There are already such calculations for the elastic $e-p$ \cite{AfanasevRadCorr,elradgen}, however, carried out within the ultrarelativistic approximation where the electron mass squared has been neglected ($m^2_{ee} \ll Q^{2}$). The code called MASCARAD \cite{AfanasevRadCorr} was 
developed for RC calculations, and another one called ELRADGEN \cite{elradgen} was developed to generate radiative events for a full Monte Carlo simulation of the PRad-type experiment. The M{\o}ller RC (events) have been calculated (generated) using the codes called MERA \cite{Ilyi05} and MERADGEN \cite{MERADGEN}. In this case the ultrarelativistic approximation was also utilized. The explicit expressions without this approximation for one-loop (i.e. vertex, self-energies and two photon exchange) contributions to M{\o}ller scatterings are presented in \cite{Kaiser:2010}, nevertheless, the contribution from hard photon emission was not considered. This contribution was taken into account in \cite{Epstein:2016lpm} where they have extended the results of \cite{Kaiser:2010} with exact single hard-photon bremsstrahlung calculations\footnote{The calculations in \cite{Epstein:2016lpm}, containing no ultrarelativistic approximation, permit a complete analysis of the next-to-leading-order (NLO) RC for both M{\o}ller and Bhabha scattering 
in the low energy kinematics of the OLYMPUS experiment.}.

For the radiative effects of the elastic $e-p$ and M{\o}ller $e-e$ scatterings that happened in the actual PRad experiment, separate event 
generators \cite{Akushevich:2015toa,PRadAnalyzer} were built, which included the NLO contributions to the Born cross sections of these scattering
processes. Ref.\,\cite{Akushevich:2015toa} has a complete set of analytical expressions for calculated RC diagrams to $e-p$ and M{\o}ller 
scatterings\footnote{The calculations of \cite{Akushevich:2015toa} do not include two-photon exchange, radiation off proton and up-down 
interference, and hadronic vacuum polarization.},
obtained within a covariant formalism and beyond the ultrarelativistic approximation, before those were calculated in \cite{Epstein:2016lpm}.
Another independent elastic $e-p$ event generator \cite{Gramolin:2014pva} was used as a cross-check. The corrections to the proton line, which 
were often neglected, were included in this generator. However, these corrections are highly suppressed due to proton's heavy mass, and are negligible 
in the PRad kinematic range. The two $e-p$ event generators were found to be in excellent agreement with each other. They also included the 
contribution from the two-photon exchange processes \cite{Tomalak:2018ere,Tomalak:2015aoa,Tomalak:2014sva}, which were estimated to be less 
than 0.2\% for the $e-p$ elastic scattering cross section in the PRad kinematic range.
All the generators are able to generate hard radiated photons, beyond the peaking approximation, by which the radiated photon will be co-linear with the
electron. This is crucial for calorimeter simulations, as the HyCal will integrate some of the radiated photons into an electron cluster, if they are close 
enough to each other when they hit the HyCal. Details and results on the NLO RC for the elastic {\em e-p} and M{\o}ller scatterings for the PRad experiment can be found in~\cite{Akushevich:2015toa}.

\begin{figure}[!ht]
\centerline{
\includegraphics[width=0.7\textwidth]{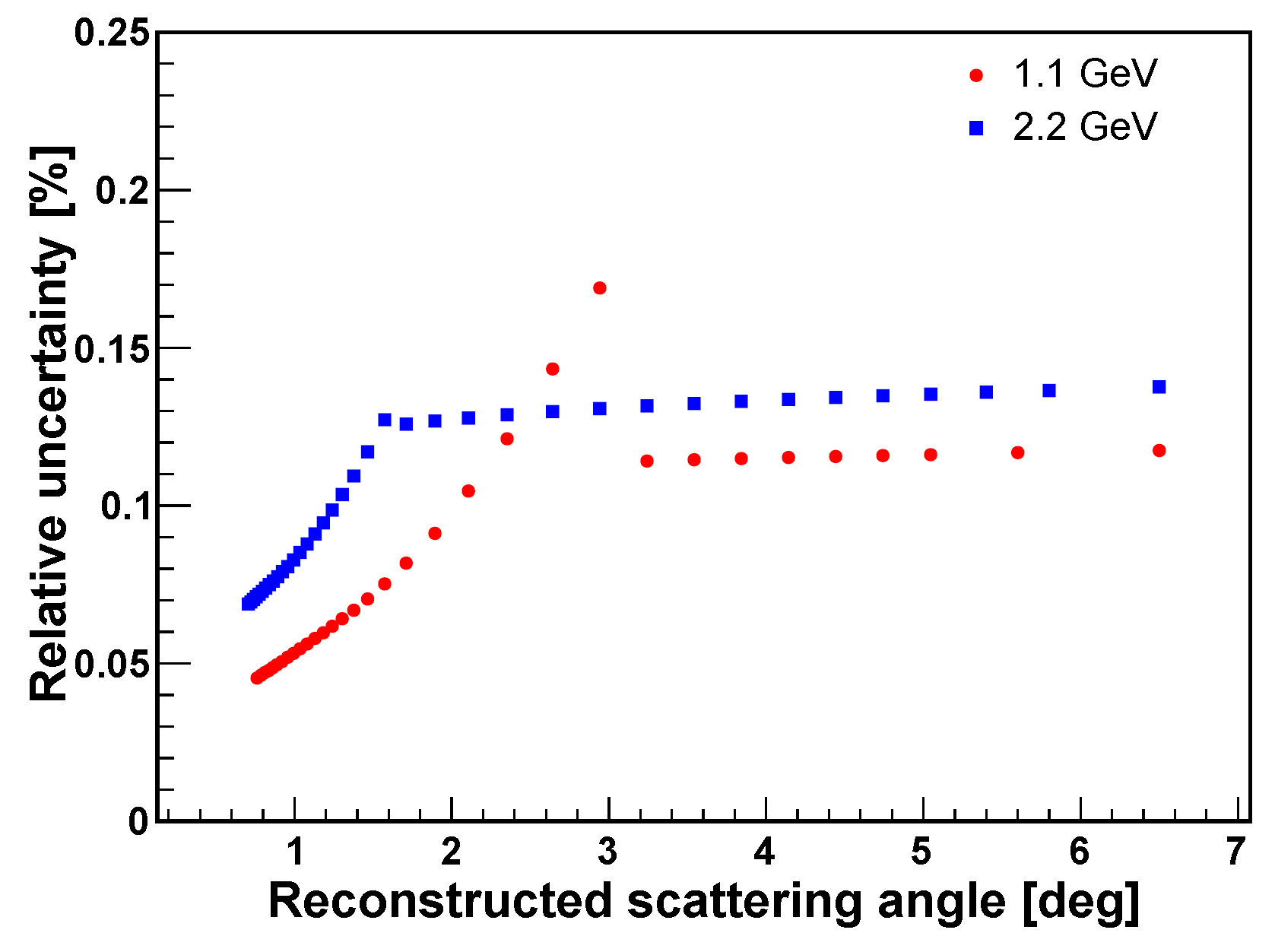}}
\caption{Relative systematic uncertainties for the cross sections due to radiative corrections for the $e-p$ and $e-e$ scatterings. The blue squares are for the 2.2~GeV energy setting, the red dots are for the 1.1~GeV energy setting. The figure is from Ref.\,\cite{Xiong:2020}.}
\label{fig:NLO_ep_ee}
\end{figure}

We would like to discuss our estimation of higher order RC systematic uncertainties based upon elastic $e-p$ and $e-e$ scatterings for PRad. If we consider both elastic $e-p$ and M{\o}ller $e-e$ scatterings, then in these processes the systematic uncertainties due to radiative corrections arise mainly from their higher order contributions to the cross sections. 
As we discussed, the NLO RC diagrams 
are meticulously worked out beyond the ultrarelativistic approximation in Ref.\,\cite{Akushevich:2015toa,PRadAnalyzer}. And these corrections also include multi-photon emission and multi-loop processes, which are approximated at the $Q^{2} \rightarrow 0$ limit by an exponentiation procedure described in \cite{Akushevich:2015toa}. Nonetheless, these higher order contributions are not calculated exactly, and the possible systematic uncertainties have been estimated by the Duke group based on the approach of Arbuzov and Kopylova \cite{Arbuzov:2015vba} developed for some of higher order RC. The estimated systematic uncertainties for both $e-p$ and M{\o}ller are correlated and $Q^{2}$-dependent. These uncertainties on the cross sections are shown in Fig.\,\ref{fig:NLO_ep_ee} for the 1.1~GeV and 2.2~GeV data sets. The $Q^{2}$-dependence is larger for the M{\o}ller RC and it affects the cross section results through the use of the bin-by-bin method \cite{Xiong:2020}. This can be seen from the uncertainties below 1.6$^{\circ}$ for the 2.2~GeV data set and below 3.0$^{\circ}$ for the 1.1~GeV data set, where the bin-by-bin method is applied. On the other hand, the $Q^{2}$-dependence for the $e-p$ RC is estimated to be much smaller relatively. 
If we transform these cross section uncertainties into the uncertainties on the proton radius, then for $e-p$ we have $\sim$\,0.0020~fm, and for M{\o}ller $e-e$ we have $\sim$\,0.0065~fm, such that the total systematic uncertainty due to NLO RC is equal to $\Delta r_{p}$ = 0.0069~fm. 

The $Q^{2}$-dependent systematic uncertainties from the M{\o}ller scattering can be suppressed by using the integrated M{\o}ller method for all angular bins, which will turn all systematic uncertainties from the M{\o}ller into normalization uncertainties for the cross sections. However, this procedure requires high precision GEM efficiency measurements particularly for the forward angular region, which cannot be achieved with the PRad setup, but can be achieved with an additional GEM plane for the PRad-II setup. 

Given that the $Q^{2}$-dependent systematic uncertainty is much larger for the M{\o}ller scattering and the potential impact on $r_{p}$ can be more significant, another independent estimate is performed by the Duke group. This estimation follows the method developed for the MOLLER experiment at JLab \cite{Aleksejevs:2013}, where the authors have calculated two-loop electroweak corrections to the parity-violating polarization asymmetry in the M{\o}ller scattering in MOLLER kinematic range. Based on their mathematical framework, we were able to estimate the contribution from the next-to-next-leading order (NNLO) diagrams on the Born cross section in the PRad kinematic range. The estimated $Q^{2}$-dependent systematic 
uncertainties are smaller than those estimated in the first approach, for any reasonable photon energy cut for the PRad experiment (from 20~MeV to~70 MeV)\footnote{In our estimation one caveat is that we estimated the NNLO RC based on a restricted set of diagrams considered in
\cite{Aleksejevs:2013}.}. Thus, we still use the uncertainty ($\Delta r_{p}$ = 0.0069~fm) from the first approach as a conservative estimate on $r_{p}$.

Next we discuss the RC systematic uncertainty in M{\o}ller scattering for PRad-II setup based on the integrated M{\o}ller method.
The common systematic uncertainty of the PRad $r_{p}$ result from \cite{Xiong:2019} is dominated by the $Q^{2}$-dependent uncertainties. In particular, it is dominated by those uncertainties that primarily affect the low $Q^{2}$ data points, such as those stemming from the M{\o}ller 
scattering. These uncertainties include the M{\o}ller RC, M{\o}ller event selection, beam energy, detector positions, etc. They are introduced into the cross section measurements by the use of the bin-by-bin method, in which one obtains the $e-p$ to $e-e$ ratio by taking the $e-p$ and $e-e$ counts from the same angular bin. In other words, the $e-p$ count in each angular bin gets a different normalization factor from the M{\o}ller $e-e$ count. 

\begin{figure}[tbp]
\begin{center}
\includegraphics[width=14cm]{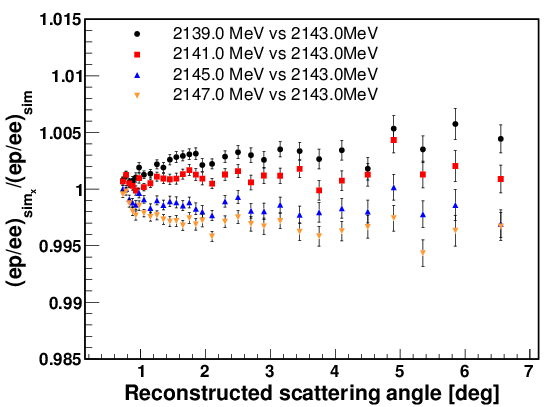}
\includegraphics[width=14cm]{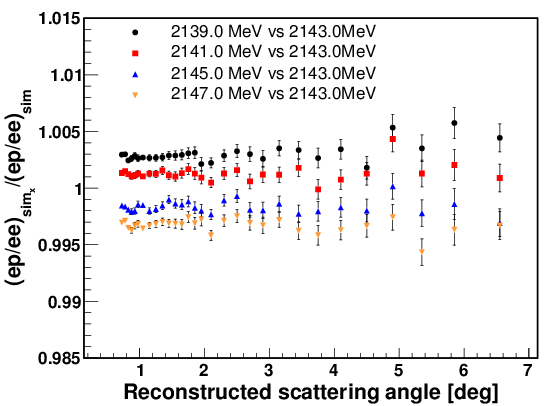}
\caption{The $e-p$ to $e-e$ ratios from simulations with different beam energies (labeled as sim$_{x}$) are plotted relative to those obtained with the nominal beam energy (labeled as sim), for the 2.2~GeV setting. In the upper plot the integrated M{\o}ller method is applied for all angular bins above 1.6$^{\circ}$. In the lower plot the integrated M{\o}ller method is applied for all angular bins.}
\label{example}
\end{center}
\end{figure}

On the other hand, the $r_{p}$ result is insensitive to the normalization uncertainties, which may shift all data points up or down at the same time. The $Q^{2}$-dependent systematic uncertainties on $r_{p}$ can be eliminated by introducing a floating parameter in the radius extracting fitter. The studies in \cite{Yan:2018bez} have already shown that the effect on $r_{p}$ is nearly zero, even with a normalization uncertainty that is as larger as 5\% (ten times larger than the typical normalization uncertainties for PRad). Thus, in order to reduce the systematic uncertainties on $r_{p}$, one can rely more on the integrated M{\o}ller method rather than on the bin-by-bin method. In this case, one would integrate the M{\o}ller counts in a fixed angular range, and use it as a common normalization factor to the $e-p$ counts from all angular bins. This will turn all systematic uncertainties from the M{\o}ller into normalization uncertainties on the cross section, and thus completely eliminate any possible effect 
on $r_{p}$. An example is illustrated in Fig.~\ref{example}, where the $e-p$ to $e-e$ ratios from simulations with different beam energies are plotted relative to those obtained with the nominal beam energy. For the upper plot, the results with scattering angles less than 1.6$^{\circ}$ are obtained with the bin-by-bin method, while the results with larger scattering angles are obtained with the integrated M{\o}ller method. There is a clear $Q^{2}$-dependent systematic uncertainty caused by the bin-by-bin method in the forward angular region. On the other hand, for the bottom plot the integrated M{\o}ller method is applied for all angular ranges. In this case, the beam energy affects mostly just the normalization of the data points. The effect on the extracted $r_{p}$ will be significantly smaller. 

While the integrated M{\o}ller method is excellent in eliminating systematic effects on $r_{p}$ due to the M{\o}ller, one would need to correct for the GEM efficiency as well, which can be cancelled by using the bin-by-bin method. This is the reason why the integrated M{\o}ller method has not been applied for the full angular range in the PRad case, since the GEM efficiency was very difficult to measure precisely in the forward angular region. This is mostly due to the HyCal finite resolution effect. In the case of PRad there was only effectively a single GEM plane. When measuring the GEM efficiency, the incident angle of the electron was measured by HyCal, the position resolution of which (on the order of 1~mm or worse) was not good 
enough to resolve various dead areas on the GEM detectors (such as those caused by the GEM spacers). In PRad-II, there will be a second GEM plane), so one can apply the integrated M{\o}ller method for the entire kinematic region. 

Thereby, the procedure described above will be applicable to PRad-II experiment that will give us almost zero systematic uncertainty on $r_{p}$, in particular for the M{\o}ller RC, however, it would be very relevant to obtain it also from the theory side.
One of our priority goals is to calculate exactly the NLO and NNLO RC in unpolarized elastic $e-p$ and M{\o}ller $e-e$ scatterings beyond
ultrarelativistic limit, when the electron mass will be taken into account at PRad/PRad-II beam energies. In this case we will have the $e-p$ and M{\o}ller radiatively corrected cross sections with both NLO and NNLO RC included. Based upon such new calculations we will also modify the event generator of \cite{PRadAnalyzer}, which has been used in the analysis of the PRad data. Its new version will be used in the analysis of the PRad-II data.

It will be an outstanding problem to calculate the corresponding one-loop and two-loop Feynman diagrams systematically. In general, it is highly desirable to develop methods for numerical semi-analytic evaluation of such diagram functions, like Feynman integrals. The problem of studying these integrals is a classic one, on which many papers have been written. However, some very basic questions still remain unanswered. For example, even in the one-loop case the precise representation of fundamental group of the base by a multi-valued function defined by a Feynman integral is unknown \cite{Bern:2013pya}. There has been tremendous number of research works accomplished on supersymmetric amplitudes on mass shell, with one of the landmark papers being Ref.\,\cite{ArkaniHamed:2012nw}. However, it is known that not all amplitudes evaluate to polylogarithms, therefore the subject of elliptic polylogarithms is being intensely studied \cite{Adams:2018ulb}. On the mathematical side, the structures of flat bundles defined by the Gauss-Manin connection are actively studied \cite{Takayuki:2017}. For generic values of parameters, it is known that the Gamma-series are a known tool to construct convergent expansions \cite{Saito:Springer2000}.

However, despite this great progress, these techniques have not been applied to the problem at hand, namely on-shell amplitudes relevant to $e-p$ or $e-e$ scatterings. One of the difficulties stems from the fact that these amplitudes need to be evaluated on the mass shell, and thus they are infrared divergent. Also, one needs to have a systematic mapping of the space of kinematic invariants and convergent expansions in a covering of this space by open cylinder domains. Besides, there is a need for a new method to expand dimensionally regulated integrals away from singularties, as well as obtain the asymptotic expansion near the singular locus. Our method is based on identification of small parameters in the corresponding domain, and expanding the integrand into series that are convergent on the chain of integration. The calculated results, namely amplitudes or cross sections, can be represented as power series, for the coefficients of which recursive relations in mathematical literature are available. Infrared regulators will be represented by off-shellness of lines and show up as overall factors. 

There is a plan to calculate the NLO and NNLO RC in $e-p$ and M{\o}ller scattering processes beyond ultrarelativistic limit. These calculations will be based upon a new method (that will address the aforementioned issues), which is under development \cite{Srednyak:2020}\footnote{The current status of the method will be reported in CFNS Ad-Hoc workshop ``Radiative Corrections",  July 9-10 (2020) at Stony Brook University, NY.}. New results on $e-e$ and $e-p$ NLO RC, which will be coming from such a new and independent method, shall be compared with the corresponding results from \cite{Akushevich:2015toa}, in order to make sure in robustness of the method before proceeding to calculations of NNLO RC contributions to the cross sections of both processes. 
In \cite{Akushevich:2015toa} such calculations have been performed for a very small scattering angle range of PRad, in $0.8^{\circ} \leq \theta \leq 3.8^{\circ}$, which corresponds to the $Q^{2}$ range of $2 \cdot 10^{-4}~{\rm GeV}^{2} \leq Q^{2} \leq 2 \cdot 10^{-2}~{\rm GeV}^{2}$. For PRad-II the planned calculations will be carried out with the lowest $Q^{2}$ at $\sim 10^{-5}~{\rm GeV}^{2}$ (corresponding to a scattering angle at $\sim 0.5^{\circ}$) up to $Q^{2}$ at $6 \cdot10^{-2}~{\rm GeV}^{2}$.


\subsection{A Comprehensive Simulation}
A comprehensive Monte Carlo simulation of the PRad setup was developed using the Geant4 toolkit~\cite{geant4}. This simulation takes into account realistic geometry of the experimental setup, and detector resolutions. 
The simulation consists of two separate event generators built for the $e-p$ and $e-e$  processes~\cite{Akushevich:2015toa, Gramolin:2014pva}. Inelastic $e-p$ scattering events were also included in the  simulation using a fit~\cite{Christy:2007ve} to the $e-p$ inelastic world data. The simulation included signal digitization and photon propagation which were critical for the precise reconstruction of the position and energy of each event in the \mbox{HyCal}.
\begin{figure}[hbtp!]
\centerline{\includegraphics[width=0.5\textwidth]{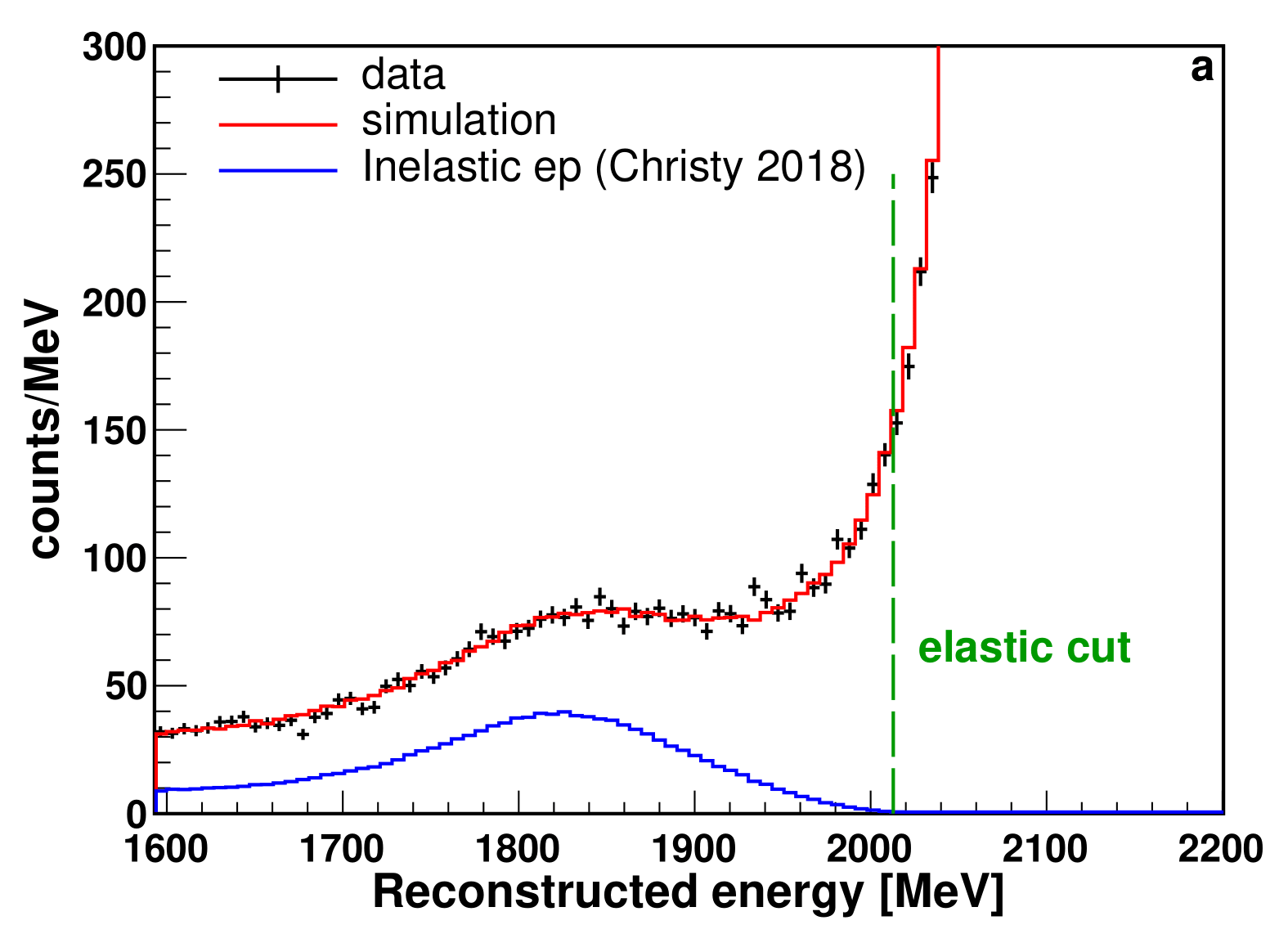}\hspace{2ex}
\includegraphics[width=0.5\textwidth]{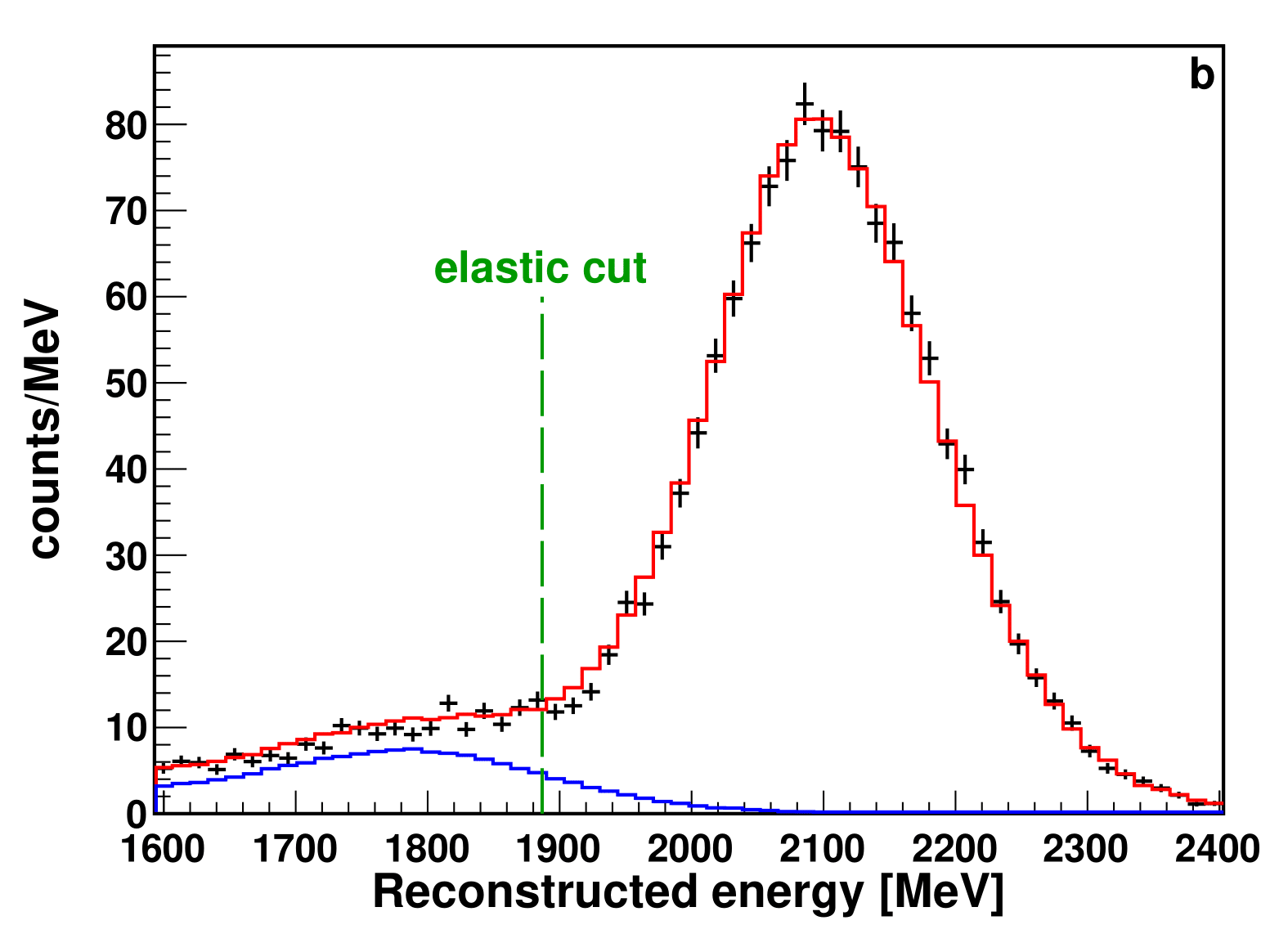}}
\caption{Comparison between reconstructed energy spectrum from the 2.2 GeV data (black) and simulation (red) for: \textbf{(a)} the PbWO$_{4}$ modules  which cover scattering angles from 3.0$^{\circ}$ to 3.3$^{\circ}$, corresponding to \qsq~around 0.014 (GeV/c)$^{2}$;   ~\textbf{(b)} the Pb-glass modules which cover scattering angles from 6.0$^{\circ}$ to 7.0$^{\circ}$, corresponding to \qsq~around 0.059 (GeV/c)$^{2}$ (largest \qsq~for PRad). Blue histograms show the inelastic $e-p$ contribution from the simulation. The green dash lines indicate the minimum elastic cut for selecting $e-p$ event for the two different detector modules. Due to the large difference in amplitudes, the elastic $e-p$ peak (amplitude ~2800 counts/MeV) is not shown in \textbf{(a)}, to display the $\Delta$-resonance peak.}
\label{fig:inel_ep}
\end{figure}
For the PRad analysis, the comprehensive Monte Carlo simulation played a critical role in the extraction of the next-to-leading order $e-p$ elastic cross section from the experimental yield. 
The simulation consists of two separate event generators built for the $e-p$ and $e-e$ processes, and they include next-to-leading order contributions to the cross section (radiative corrections), such as Bremsstrahlung, vacuum polarization, self-energy and vertex corrections.
The calculations of the $e-p$ elastic and M{\o}ller radiative corrections are performed 
within a covariant formalism, without the usual ultra relativistic approximation~\cite{Akushevich:2015toa}, where the mass of the electron is neglected. The two generators also include contributions from two-photon exchange processes~\cite{Tomalak:2018ere, Tomalak:2015aoa, Tomalak:2014sva}. A second independent $e-p$ elastic event generator~\cite{Gramolin:2014pva} was used as a cross check.  The radiative corrections to the proton, which are typically neglected, were included in this generator. The two $e-p$ generators were found to be in excellent agreement. 
   
Inelastic $e-p$ scattering events were included in the simulation using an empirical fit~\cite{Christy:2007ve} to the $e-p$ inelastic scattering world data. Inelastic $e-p$ scattering contributes a background to the $e-p$ elastic spectrum which, when included in the simulation was able to reproduce the measured elastic $e-p$ spectrum as shown in Fig.~\ref{fig:inel_ep}. 
In the PbWO$_{4}$ segment of the calorimeter, there was a clear separation between the elastic and inelastic $e-p$ events, and it was established that the position and amplitude of the $\Delta$-resonance peak in the simulation agreed with the data to better than 0.5\% and 10\%, respectively. The  $\Delta$-resonance contribution was found to be negligible ($\ll 0.1\%$) for the PbWO$_{4}$ segment of the HyCal, and no more than 0.2\% and 2\% for the Pb-glass segment, at 1.1~GeV and 2.2~GeV, respectively. 
The generated scattering events were propagated within the Geant4 simulation package, which included the detector geometry and materials of the PRad setup. This enabled a proper accounting of the external Bremsstrahlung of particles passing through various materials along its path. The simulation included photon propagation and digitization of the simulated events. These steps were critical for the precise reconstruction of the position and energy of each event in the \mbox{HyCal}.

To simuate the proposed PRad-II experiment, the comprehensive simulation of the PRad experiment was updated to include the second plane of GEM detectors and the scintillator detector. 
The comprehensive simulation was used to generate mock data for the PRad-II experiment. The mock data was then used with the PRad analysis package to extract the cross section and form factor $G_{E}^{p}(Q^2)$. The robust $r_p$ extraction method developed for PRad (described in sec.~\ref{sec:PRad}) was use to obtain $r_p$ from the mock data (shown in Fig.~\ref{fig:PRad2_ge}). The simulations was also used for estimating the expected rates, the systematic uncertainties and the projected results of the PRad-II experiment (see sec.~\ref{sec:errors}).
\begin{figure}[hbtp!]
\centerline{\includegraphics[width=0.45\textwidth]{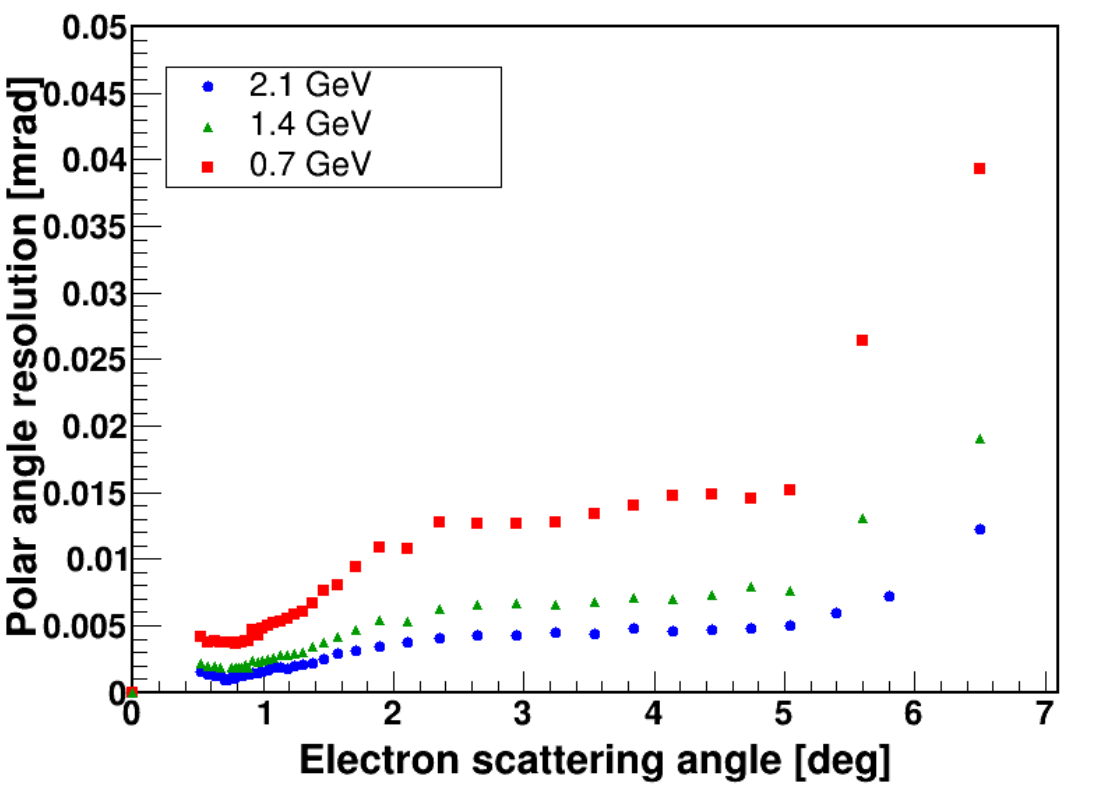}\includegraphics[width=0.45\textwidth]{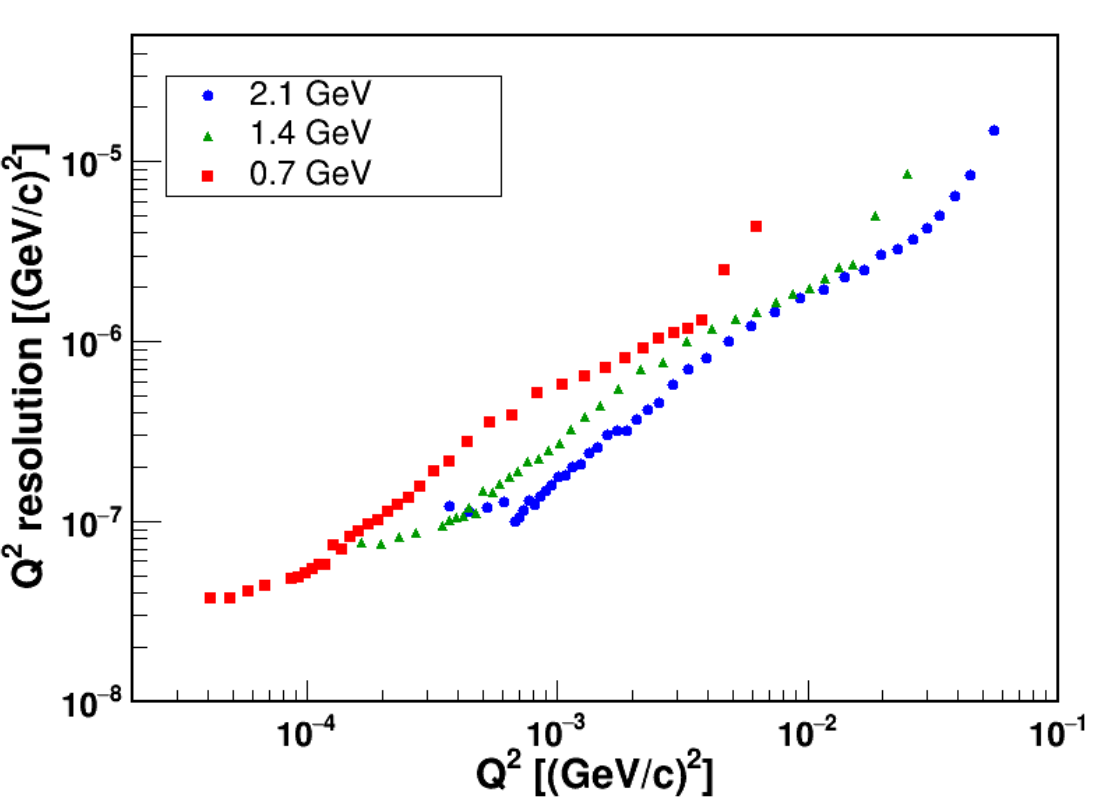}}
\caption{The anglular resolution (left) over the angular range covered in the experiment and $Q^2$ resolution (right) as a function of $Q^2$, at 0.7~GeV~(red), 1.4~GeV (green) and 2.1~GeV~(blue) electron beam energy.}
\label{fig:q2res}
\end{figure}
The  simulation as also used to show that two layers of   coordinate detectors will provide an angular resolution of 0.001 - 0.004 mrad for the smallest angle at the 0.7 - 2.1 GeV  beam energy and 0.004 - 0.04 mrad for the largest angle covered at these beam energies. The energy and angular resolutions were used to obtain the $Q^2$ resolution shown in Fig.~\ref{fig:q2res}. The angular resolution is used to determine the size of the $Q^2$ bins used to extract the cross section and electric form factor from the simulated yield.

\subsection{Rates and beamtime request}

The expected rates are calculated assuming the hydrogen gas-flow target used in the PRad experiment, which achieved an areal density of  $2 \times 10^{18}$ H atoms/cm$^{2}$. We propose to run the PRad-II experiment with three different beam energy settings, 0.7~GeV, 1.4~GeV and 2.1~GeV.
The projected scattering angle coverage is from 0.50$^{\circ}$ to 7.00$^{\circ}$ for all energy settings. All scattering angles below 5.2$^{\circ}$ are expected to have a full azimuthal angular coverage so that the geometric acceptance factor $\epsilon_{geom}$ is nearly 1. Larger scattering angles are covered by the corners of HyCal and thus, only part of the azimuthal angles are covered. In the worst case (6.0$^{\circ}$ to 7.0$^{\circ}$), the geometric acceptance factor can drop down to about 0.15. For estimating the overall rate, the acceptance is still close to 1 as the $e-p$ cross section falls roughly as $1/\rm{sin}(\theta/2)^{4}$. The detector efficiency $\epsilon_{det}$ will be dominated by the GEM efficiencies, which is about 93\% for the PRad GEMs. For PRad-II, one shall require two coincident hits on the two GEM planes for each scattered electron. This lead to about 86\% for $e-p$ events and about 75\% for $e-e$ events.

The event rate can be estimated using:
\begin{equation}
    N = N_{e} \cdot N_{tgt} \cdot \Delta \sigma \cdot \epsilon_{geom} \cdot \epsilon_{det}
\end{equation}
where $N_{e}$ is the number of incident electron per second, $N_{tgt}$ is the target areal density, $\Delta \sigma$ is the integrated elastic cross section, for which we will use the Born level cross section for simplicity. This leads to $\Delta \sigma$ of 6.940$\times 10^{-27}$~cm$^{2}$, 1.730$\times 10^{-27}$~cm$^{2}$ and 0.766$\times 10^{-27}$~cm$^{2}$ for the 0.7, 1.4 and 2.1~GeV beam energy setting, respectively, for the scattering angular ranges mentioned above. The choice of beam current is based on the expected maximum data rate allowed by the new GEM detector DAQ (25 kHz), the expected trigger rate for the calorimeter and maximum power allowed on the Hall-B Faraday cup (160 ~W). The Faraday cup is essential for the background subtraction using the empty target data.
We plan to use a current of 20~nA (1.248$\times 10^{11} e^{-}$/s) at 0.7~GeV beam energy and 70~nA (4.370$\times 10^{11} e^{-}$/s) current at both 1.4 and 2.1~GeV beam energies. The 70~nA current limit is imposed by the maximum power allowed on the Hall-B Faraday cup. 

For the $e-e$ scattering, if we require double-arm M{\o}ller detection, the scattering angular coverage will be 0.5$^{\circ}$ to 9.5$^{\circ}$ for the 0.7~GeV (the electron at scattering angles larger than HyCal acceptance will be detected by the proposed scintillating detector) and 0.5$^{\circ}$ to 4.8$^{\circ}$ for the 1.4~GeV and 0.5$^{\circ}$ to 3.2$^{\circ}$ for the 2.1~GeV. In this case, the detector efficiency will be 0.75 as we requires two hits on the two separated GEM planes for both scattered electrons. The event rates for $e-p$ scattering and $e-e$ scattering are shown in Table.~\ref{table:request_beam_time}.


We are requesting 4 days of beam time for 0.7~GeV, 5 days for 1.4~GeV and 15 days for 2.1~GeV production runs. For all energy settings, these will ensure that the statistical uncertainty of the largest angular bin (6.0$^{\circ}$ to 7.0$^{\circ}$) to be about 0.3\%, which is about 3 times smaller than that for the PRad experiment. We are also requesting an additional 33\% of beam time (8 days) for various empty target measurements, for the purpose of the empty target subtraction and beam background studies. The total requested beam time for various stage of the proposed experiment is listed in Table.~\ref{table:request_beam_time}

\begin{table}[hbt!]
\begin{center}
\begin{tabular}{ | l | c | c | c |}
\hline
 item & $e-p$ event rate &$e-e$ event rate &Time \\\hline
      & M e$^-$/day & M e$^-$/day & days\\
\hline
Setup checkout, tests and calibration & & & 7.0 \\
\hline
Production at 0.7~GeV &129 &230 & 4.0 \\
\hline
Production at 1.4~GeV &112 & 205& 5.0 \\
\hline
Production at 2.1~GeV &50 &90 & 15.0 \\
\hline
Empty target runs & & & 8.0 \\
\hline
Energy change & & & 1.0 \\
\hline
Total & & & 40.0 \\
\hline
\end{tabular}
\caption{The PRad-II event rate and beam time request}
\label{table:request_beam_time}
\end{center}
\end{table}

\subsection{Robust extraction of the proton charge radius}

\subsubsection{Method and main results for PRad}
\label{sec:PRad}
There are various well-developed proton electric form factor, $G_{E}$, models, such as \cite{Kelly:2004hm,arrington2004implications,arrington2007precise,ye2018proton,alarcon2017nucleon,alarcon2018nucleon,alarcon2018accurate,bernauer2014electric}. Most of them have been fitted with experimental data in high \(Q^2\) ranges. Meanwhile, these models have different kinds of extrapolation in lower \(Q^2\) ranges, for example, in the PRad \(Q^2\) range, which is from 
\(2 \cdot 10^{-4}\ {\rm to}\ 2 \cdot 10^{-2}~{\rm (GeV/c)^2}\). Such studies have been accomplished in Ref.~\cite{Yan:2018bez}, which in particular gives a general framework with input form factor functions and various fitting functions (fitters) for determining functional forms that allow for a robust extraction of the input charge radius of the proton, $R_{p}$, for the PRad experiment. 

The robustness of any suitable fitter when extracting the root-mean-square (RMS) charge radius of the proton in a lower \(Q^2\) range can be tested by fitting pseudo-data generated in that range by different $G_{E}$ models \cite{Yan:2018bez}. In the fitting procedure, depending on a fitting function, different bias and variance are obtained. The bias is calculated by taking the difference between the fitted radius mean value and the input radius value from a model: 
${\rm bias} \equiv \Delta R_{p}{\rm [bias]} = R_{p}{\rm [mean~fit]} - R_{p}{\rm [input]}$. The variance is the fitting uncertainty ($\sigma$) represented by the RMS value of a fitting result. To control the total uncertainty, the number of free parameters in a fitting function should not be too large. Otherwise, the variance from the fitting will be very large. If the variance coming out from a given fit is small and the bias is within this variance, then the corresponding fitter is considered to be robust (the figures in this note show it quantitatively). To compare the goodness between different robust fitters, the quantity called
root-mean-square-error (RMSE) is used:
\begin{equation}\label{rmse}
{\rm RMSE} = \sqrt{{\rm bias}^2 + {\sigma}^2}\,.
\end{equation}
The smaller the RMSE value is, the better the corresponding fitter is.
In this section, we concisely show the method and main results from \cite{Yan:2018bez} on $R_p$'s robust extraction for PRad. In Sec.~\ref{PRad-II} we present our new results for PRad-II but using the same method of PRad's $R_p$ extraction.

\paragraph{Generators:}
Various $G_E$ generators (models) have been used in Ref.~\cite{Yan:2018bez} for generating pseudo-data in the PRad $Q^2$ range: namely, {\it Kelly-2004} \cite{Kelly:2004hm}, {\it Arrington-2004} \cite{arrington2004implications}, {\it Arrington-2007} \cite{arrington2007precise}, {\it Ye-2018} \cite{ye2018proton}, {\it Alarcon-2017} \cite{alarcon2017nucleon,alarcon2018nucleon,alarcon2018accurate}, {\it Bernauer-2014} \cite{bernauer2014electric}, as well as {\it Dipole}, {\it Monopole}, and {\it Gaussian} \cite{borkowski1975determination}.

\paragraph{Fluctuation adder and pseudo-data generation procedure:}
In the PRad experiment, there are thirty three bins from 0.7$^{\circ}$ to 6.5$^{\circ}$ at 1.1~GeV beam energy, and thirty eight bins from 0.7$^{\circ}$ to 6.5$^{\circ}$ at 2.2~GeV. 
To mimic the bin-by-bin statistical fluctuations of the data, the $G_E$ pseudo-data statistical uncertainty is smeared by adding $G_E$ (in each $Q^2$ bin) with a random number following the Gaussian distribution, $\mathcal{N}(\mu, \sigma^2_g)$, given by
\begin{equation}
\mathcal{N}(\mu,\,\sigma_{g}^{2}) = \frac{1}{\sqrt{2\pi\sigma_{g}^{2}}}\,e^{-\frac{(G_{E} - \mu)^{2}}
{2\sigma_{g}^{2}}}\,,
\label{eq:eqn_Gaus}
\end{equation}
where $\mu =0$ and $\sigma_g = \delta G_E$, and $\delta G_E$ comes from the statistical uncertainty of the PRad data. In the case of the PRad-II experiment, $\delta G_E$ in each bin will be the half of $\delta G_E$ in the PRad data, by assuming that the PRad-II statistics will have four times of that of PRad (discussed in the next section). Let us also give some more details on the pseudo-data generation and fitting procedure:
\begin{itemize}
\item[(i)] To add the statistical fluctuations to the final results, the seventy one (thirty three + thirty eight) generated pseudo-data points are added by seventy one different random numbers according to Eq.~(\ref{eq:eqn_Gaus}).

\item[(ii)] The set of pseudo-data are fitted by a specific fitter $f_E(Q^2)$. In this procedure, the pseudo-data points at 1.1 GeV and 2.2 GeV are combined and fitted by the fitter with two different floating parameters corresponding to two different energy setups. The other fitting parameters in the fitter are required to be the same for both energy setups.

\item[(iii)] The fitted radius is calculated from the fitted function in (ii), with 
\begin{equation}
R_{p}{\rm [fit]} = \left( -6 \left.\frac{\mathrm{d} f_{E} (Q^2)} {\mathrm{d}Q^2} \right|_{Q^{2}=0} 
\right)^{1/2}\,.
\end{equation}

\item[(iv)] The above steps are repeated for 10,000 times for obtaining 10,000 sets of $G_E$ pseudo-data diluted by Eq.~(\ref{eq:eqn_Gaus}), and 10,000 $R_{p}{\rm [fit]}$ values are also calculated.

\item[(v)] $R_{p}{\rm [mean~fit]}$ is the mean value of the 10,000 $R_{p}{\rm [fit]}$ results, and the variance is the RMS value of this $R_{p}{\rm [fit]}$ distribution, which is also determined.
\end{itemize}

\begin{figure}[hbt!]
\centering
\includegraphics[scale=0.45]{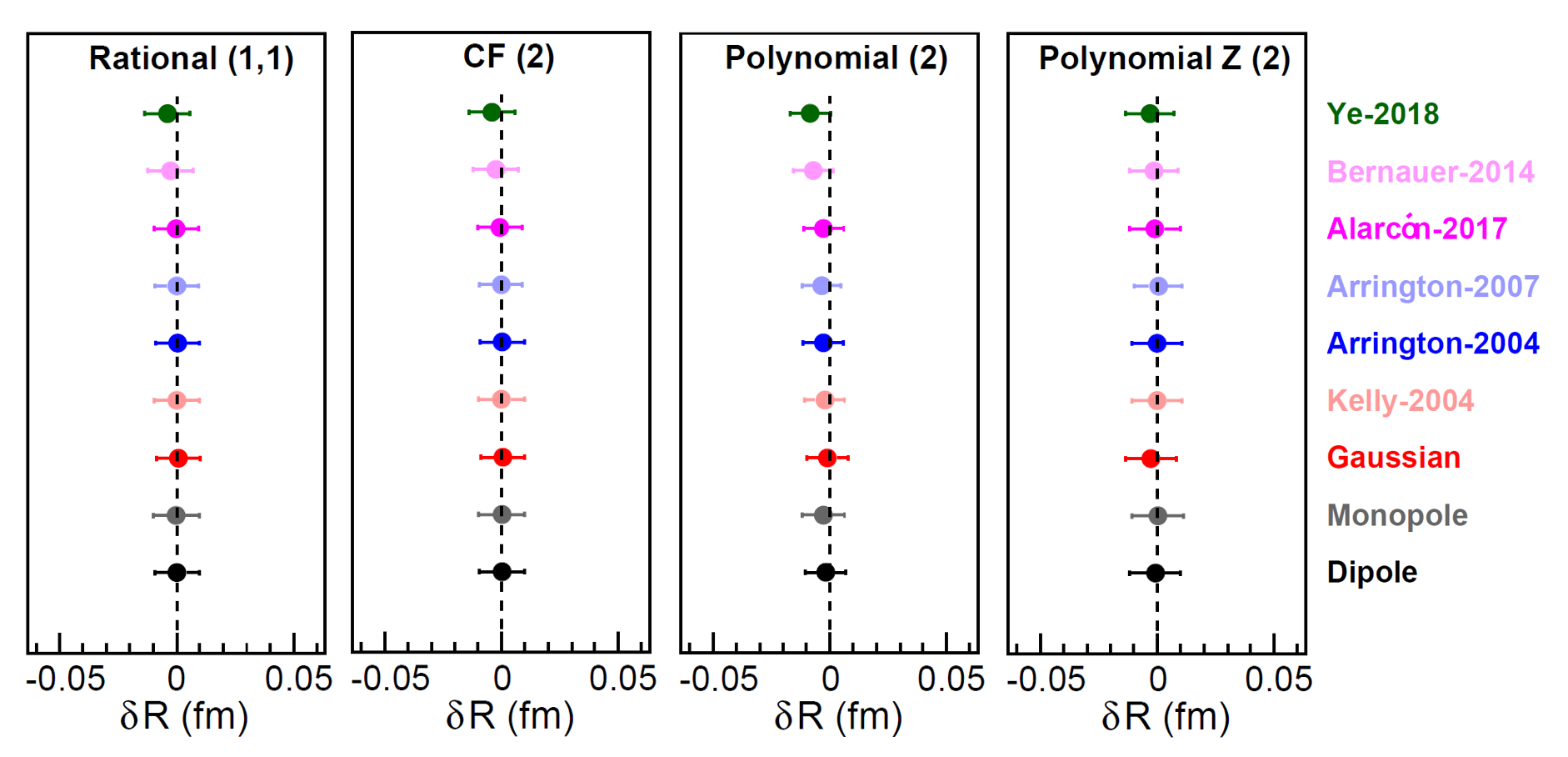}
\caption{The variance from the fitted PRad pseudo-data generated by nine $G_{E}$ models using the Rational (1,1), $2^{\rm nd}$-order CF, $2^{\rm nd}$ order polynomial expansion of $Q^2$, and $2^{\rm nd}$-order polynomial expansion of $z$, for which the bias is smaller than the variance. This figure is from \cite{Yan:2018bez}.}
\label{fig:PRad_fitters}
\end{figure}

\paragraph{Fitters:} \label{Sec:PRad_fitter}
One of the best fitters determined in Ref.~\cite{Yan:2018bez}, which robustly extracted $R_p$ for PRad, is the Rational (1,1), based on the multi-parameter rational-function, Rational (N,M) of $Q^2$, given by
\begin{equation}
f_{rational}(Q^2)=p_0 G_E(Q^2)=p_0\frac{1+\sum_{i=1}^{N} p_i^{(a)} Q^{2i}}{1+\sum_{j=1}^{M} p_j^{(b)} Q^{2j}}\,,
\end{equation}
where $p_0$ is a floating normalization parameter, and $p_i^{(a)}$ and $p_j^{(b)}$ are free fitting parameters.
For the Rational (1,1), the orders $N$ and $M$ are equal to one, and the input radius is calculated by 
$R_p = \sqrt{6\left(p_1^{(b)}-p_1^{(a)}\right)}$. Tho more robust fitters were found to be the $2^{\rm nd}$-order continuous fraction (CF) and $2^{\rm nd}$-order polynomial expansion of $z$. The other fitter functions, used to fit the generated pseudo-data in \cite{Yan:2018bez}, are the Dipole, Monopole, Gaussian, and multi-parameter polynomial expansion of $Q^2$. Although the $2^{\rm nd}$-order CF exactly has the same functional form as the Rational (1,1), in Fig.~\ref{fig:PRad_fitters} we show the results from the three best fitters plus also the 
$2^{\rm nd}$-order polynomial expansion of $Q^2$. One can see that the bias remain well within variance in the first, second and fourth plots for all the nine models, as shown in Fig.~\ref{fig:PRad_fitters}. In particular, the Rational (1,1) controls both the variance and RMSE at best. As a result, PRad used the Rational (1,1) to obtain the proton radius \cite{Xiong:2019}.
\begin{figure}[hbt!]
\centering
\includegraphics[scale=0.48]{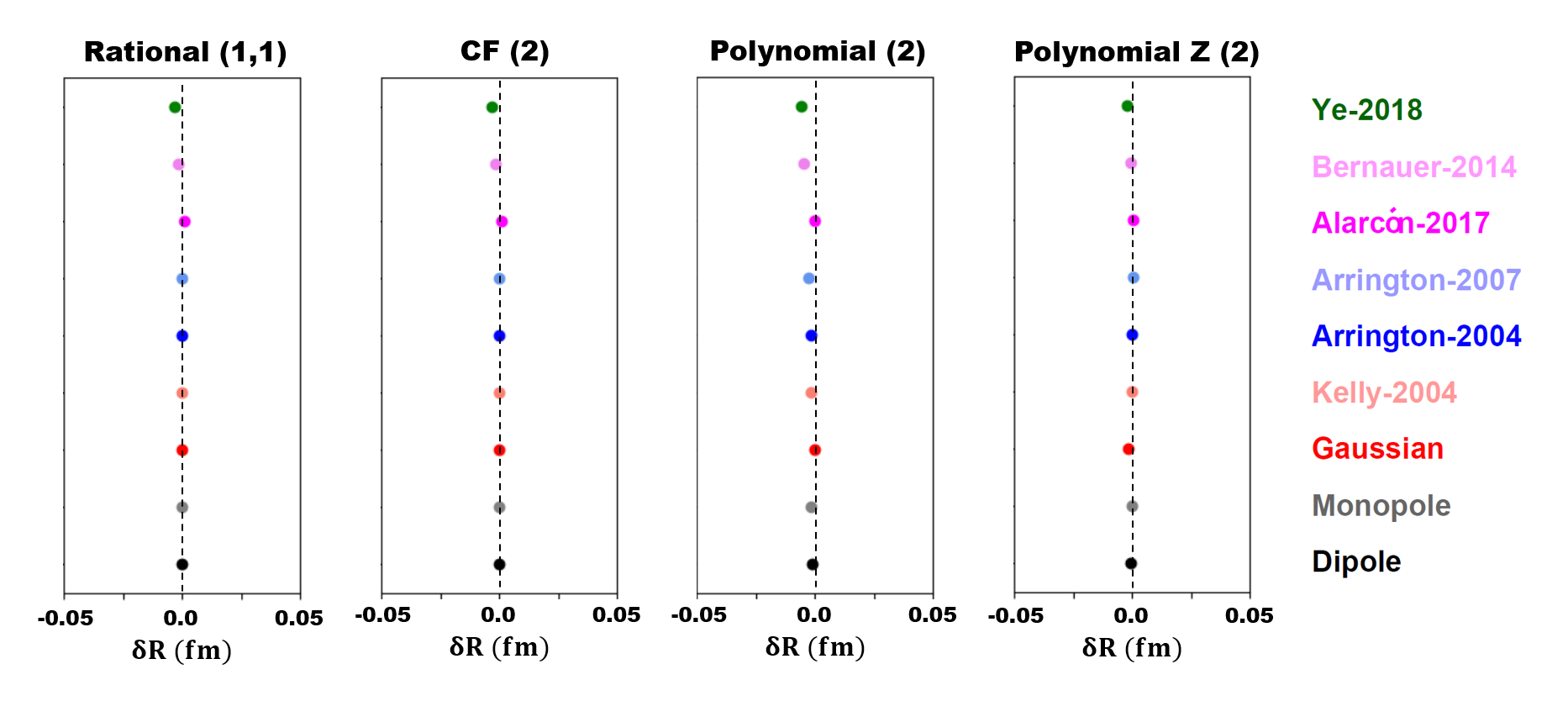}
\caption{Four variance-bias plots from fits with pseudo-data generated by the considered nine  proton $G_E$ models, made analogously to Fig.~\ref{fig:PRad_fitters}, but for the PRad-II statistics. In these plots, the error bars are too small to be seen.}
\label{fig:PRad_II_fitter}
\end{figure}
\begin{figure}[hbt!]
\centering
\includegraphics[scale=0.525]{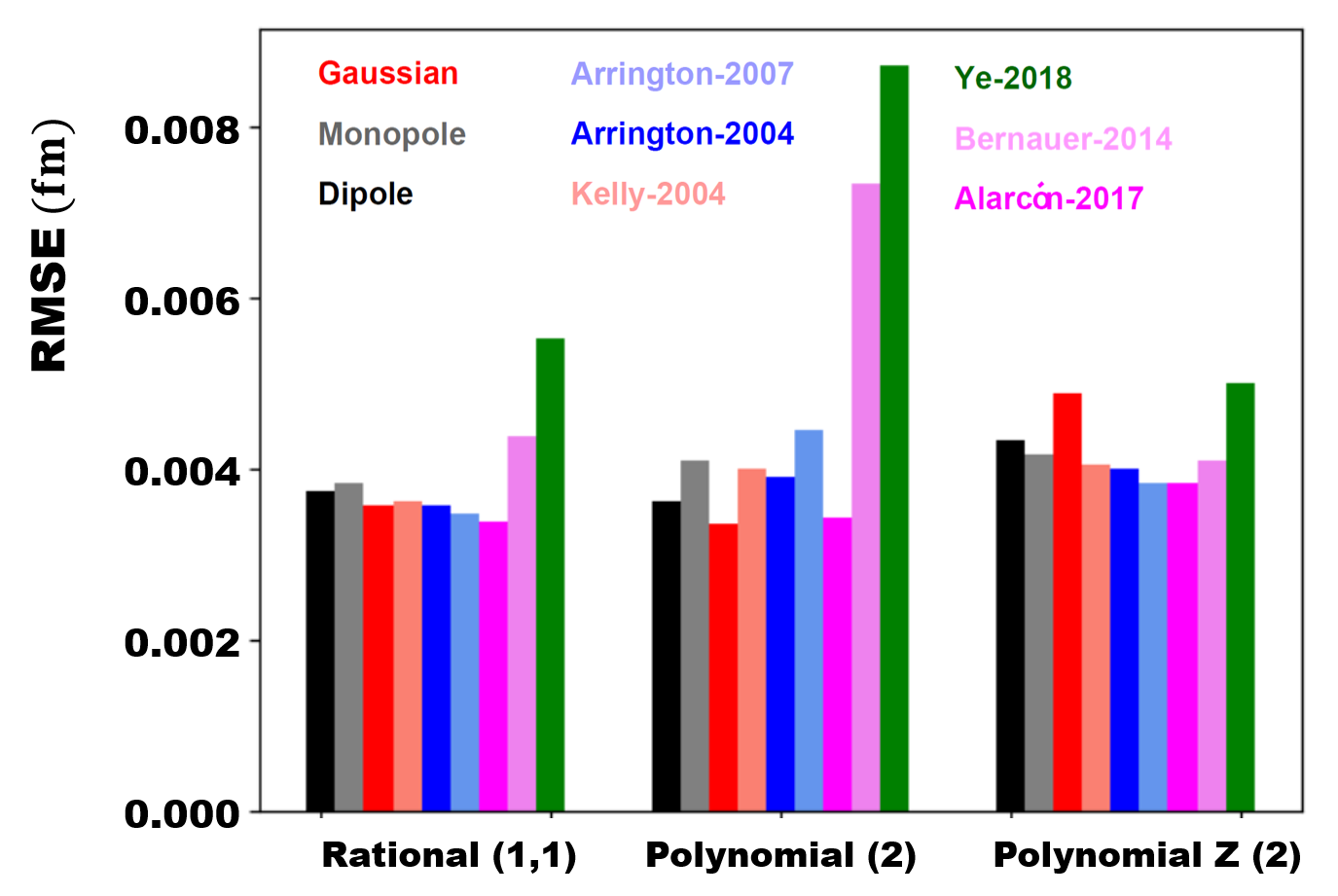}
\caption{The RMSE for PRad-II obtained from all the fitters based upon the nine different proton $G_E$ models under consideration.}
\label{fig:PRad_II_RMSE}
\end{figure}

\subsubsection{Projections for PRad-II}
\label{PRad-II}
Given the method and procedure for robustly extracting the proton radius, we can now look into PRad-II, for which the statistical uncertainty for measuring $R_p$ is planned to be $\sim$ 1/4 of that of PRad. In order to test whether the Rational (1,1) is still suitable for this case, the statistical uncertainty of $G_E$ is taken to be the half of $\delta G_E$ in Eq.~(\ref{eq:eqn_Gaus}). By using the nine different proton electric form factor models for generating 10,000 sets of $G_E$ pseudo-data, and then fitting them with the different fitters, we obtain the results shown in Fig.~\ref{fig:PRad_II_fitter}. The other fitters mentioned in Sec.~\ref{Sec:PRad_fitter} are also tested, but they are not as good as the ones shown here.

We notice that the fitters of the Rational (1,1) and $2^{\rm nd}$-order polynomial expansion of $z$ are still robust with the statistics of PRad-II. However, the latter gives a larger variance compared to that obtained from the Rational (1,1). By comparing the RMSE from Eq.~(\ref{rmse}) in Fig.~(\ref{fig:PRad_II_RMSE}), one can see that overall the Rational (1,1) has the smallest RMSE values for all nine models.

\subsubsection{Summary}
As shown in Fig.~\ref{fig:PRad_II_fitter}, the {\it Ye-2018} model gives a much larger bias compared to the other models. The bias from fitting the Rational (1,1) with the pseudo-data generated by the {\it Ye-2018} model is 0.476\%. We can consider this number as an upper bound, which corresponds to a $3\sigma$ uncertainty. Then $1\sigma$ of the bias will be 0.159\% (0.0013~fm of the PRad-II projected uncertainty in the proton radius). If we add 0.0013~fm quadratically to the total uncertainty, then its absolute increment by considering this number will be 0.0001~fm, which is a very small number.

\subsection{Estimated uncertainties and projected results}
\label{sec:errors}

The major improvement for the PRad-II $r_{p}$ result comes from the proposed use of a second GEM detector plane, which allows for more precise determination of the detector efficiency (see Fig.~\ref{fig:GEM_new_eff} in Sec.~\ref{sec:GEM}). This in turn will enable the use of integrated M{\o}ller method over the full angular range. This alone can already reduce the total systematic uncertainty by about a factor of 2, if the GEM efficiency is determined to better than 0.1\% precision. As discussed earlier in this proposal, the integrated M{\o}ller method converts the \qsq ~dependent systematic uncertainties due to M{\o}ller scattering events into normalization type uncertainties which do not contribute to the systematic uncertainties of $r_{p}$. Such systematic uncertainties include the M{\o}ller event selection, M{\o}ller radiative correction, acceptance and beam energy related uncertainties. The contribution from uncertainty in detector acceptance was determined by shifting the GEM detectors by $\pm$2~mm in the simulation, which resulted in a $\sim$ 0.0002~fm change in the extracted $r_p$ when using the integrated M{\o}ller method.
Similarly, the beam energy related uncertainty was determined by shifting the 0.7~GeV electron beam energy by $\pm$0.5 MeV (the measured uncertainty for the 1.1 GeV beam during PRad) in the PRad-II simulation which had negligible impact on the extracted $r_p$.
\begin{table}[hbt!]
\begin{center}
\begin{tabular}{ | c | c | c | c | c |}
\hline
Item &PRad $\delta r_{p}$ [fm] & PRad-II $\delta r_{p}$ [fm] &Reason \\
\hline
Stat. uncertainty &0.0075 & 0.0017 & more beam time \\\hline
GEM efficiency  &0.0042 & 0.0008& 2nd GEM detector\\
\hline
Acceptance  &0.0026 & 0.0002&2nd GEM detector\\
\hline
Beam energy related &0.0022 & 0.0002&2nd GEM detector\\
\hline
Event selection  &0.0070 & 0.0027 & 2nd GEM + HyCal upgrade\\
\hline
HyCal response  & 0.0029 & negligible & HyCal upgrade\\
\hline
  && & better vacuum\\ 
 Beam background &0.0039  & 0.0016 & 2nd halo blocker\\
& & & vertex res. (2nd GEM)\\\hline
Radiative correction &0.0069 & 0.0004&improved calc. \\
\hline
Inelastic $ep$ & 0.0009 & negligible& - \\
\hline
G$_{M}^{p}$ parameterization &0.0006 & 0.0005& HyCal upgrade\\
\hline
Total syst. uncertainty  &0.0115 & 0.0032& \\
\hline
Total uncertainty &0.0137 & 0.0036& \\
\hline
\end{tabular}
\end{center}
\caption{The uncertainty table for $r_{p}$ from the PRad experiment, and the projected uncertainties for PRad-II. Uncertainties are estimated using the Rational (1,1) function.}
\label{table:PRad2_uncertainty}
\end{table}

The reduction in the uncertainties due to event selection is a result of both the second GEM detector and the HyCal upgrade, while the uncertainty due to HyCal detector response is reduced because of the upgrade of HyCal to an all PbWO$_4$ calorimeter as shown in Fig.~\ref{fig:hycal_upgrade} in Sec.~\ref{sec:calor}.  The uncertainty from the beam-line background rejection is reduced because of the anticipated better beam-line vacuum, the additional beam halo blocker and the improved vertex reconstruction and tracking with the second GEM detector, as shown in Fig.~\ref{fig:new_z_recon} in Sec.~\ref{sec:GEM}. The proposal also includes reduced uncertainty due to radiative corrections because of the new calculations that include the next-to-next-leading order Feynman diagrams in the radiative correction (see Sec.~\ref{sec:newrad}).
The projected result also assumes a factor of $\sim$~19 increase of the total statistics compared to PRad. This leads to $>$~4 times reduction of the statistical uncertainty of $r_{p}$. The additional statistics will also slightly improve the systematic uncertainties that are statistics dependent, such as the statistical uncertainties in the detector efficiencies and calibrations.  The total systematic uncertainty is about a factor of 3.6 times smaller than that from the PRad and the total uncertainty is about 3.8 times smaller. The  projected uncertainties for PRad-II are shown in Table.~\ref{table:PRad2_uncertainty}.

\begin{figure}[hbt!]
\centerline{
\includegraphics[width=0.5\textwidth]{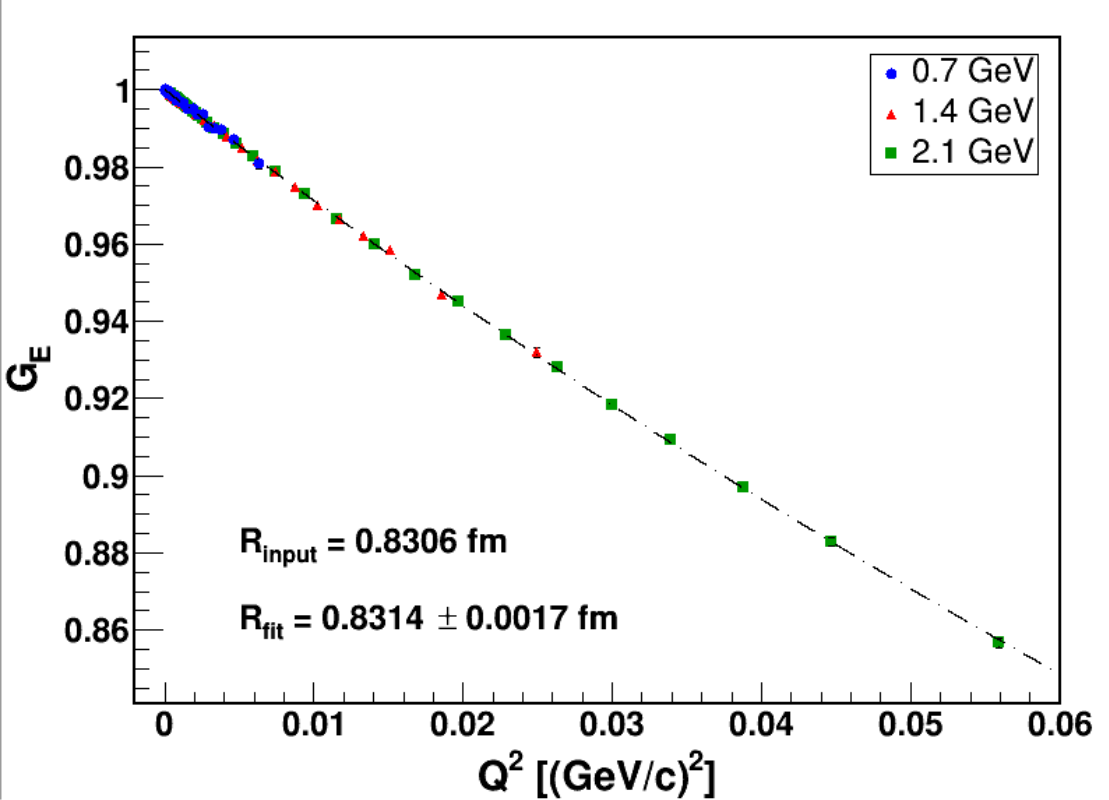}
\includegraphics[width=0.5\textwidth]{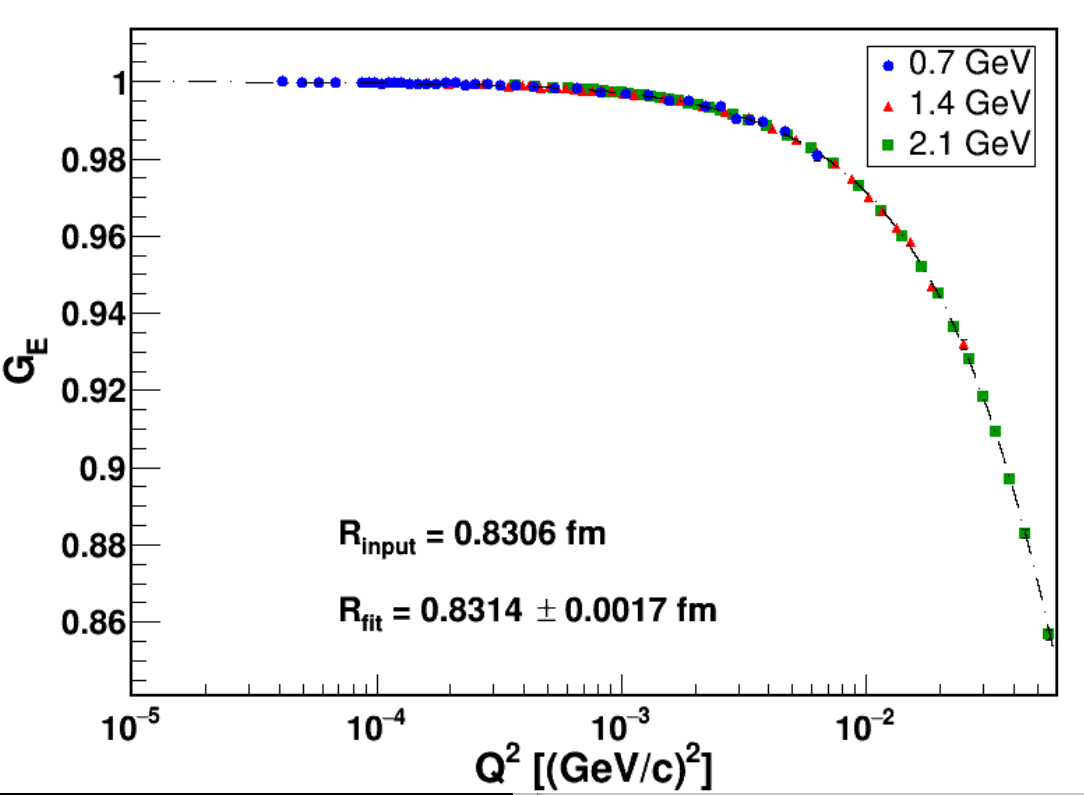}}
\caption{(left) The $G_{E}^{p}(Q^2)$ obtained from the mock data generated by the comprehensive simulation at 0.7 (blue), 1.4 (red) and 2.1 GeV (green) beam energies, which is then fit to a rational (1,1) functional form (dashed line) to extract the $r_p$. (right) The same information shown in log scale on the x-axis.}
\label{fig:PRad2_ge}
\end{figure}

\begin{figure}[hbt!]
\centerline{
\includegraphics[width=1.0\textwidth]{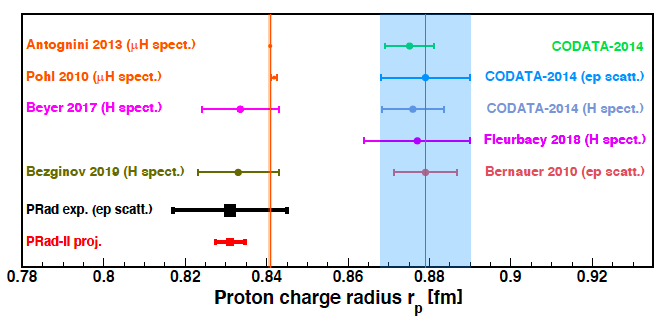}}
\caption{The projected $r_{p}$ result from PRad-II, showing along with the PRad result and other measurements.}
\label{fig:PRad2_proj}
\end{figure}

The comprehensive simulation of the PRad-II experiment was used to generate 10,000 mock data sets at the 3 proposed beam energies. The $G_{E}^{p}(Q^2)$ obtained from the mock data are shown in Fig.~\ref{fig:PRad2_ge}. The $G_{E}^{p}(Q^2)$ was fit to a rational (1,1) functional form to extract the $r_p$ as shown in Fig.~\ref{fig:PRad2_ge}.
The $r_p$ extracted from the fits along with the statistical uncertainty is $r_p =$ 0.8314 $\pm$ 0.0017 fm.\\
The projected $r_p$ from the PRad-II experiment along with other measurements and the CODATA values are shown in Fig.~\ref{fig:PRad2_proj}.

\section{Summary}
\label{sec:summary}
We propose an enhanced proton rms charge radius experiment, PRad-II, which will achieve a factor of 3.8 lower uncertainty in the extracted radius compared to PRad. This improvement in uncertainty will be achieved by; i) collecting over an order of magnitude more statistics, reducing the statistical uncertainty by a factor of 4. This is especially important for the highest Q$^2$ region covered in the experiment and for reducing the total systematic uncertainties. ii) Adding a new GEM coordinate detector to incorporate tracking capability in the experiment. This will enable using the reconstructed interaction vertex to significantly reduce the beam-line background in the experiment. This is especially important for the smallest scattering angles which is critical for reaching the lowest Q$^2$ range of 10$^{-5}$ GeV$^2$ for the first time in lepton scattering experiments. iii) Upgrade of HyCal to an all PbWO$_4$ calorimeter. This will significantly enhance the uniformity of the detector package, a critical requirement for the precise and robust extraction of the proton rms rcharge radius. iv) Upgrade of the FASTBUS-based HyCal readout electronics to a flash-ADC-based system speeding up the DAQ system by a factor of 7 and reducing the total beam time to achieve the required statistics. v) Improvements to the beamline vacuum, and a second beam halo blocker upstream of the tagger,  to further suppress the beamline background. This is critical for a clean separation of the $ep$ and $ee$ scattering events at the small scattering angles covered in the experiment. vi) Improved radiative corrections for both $ep$ and $ee$ scattering which will significantly reduces the uncertainty due to radiative corrections. 

\noindent
In addition to the factor of 3.8 reduction in the total uncertainties compared to PRad, we also propose to enhance the range of Q$^2$ covered in PRad-II. The proposed experiment will reach the lowest Q$^2$ range of 10$^{-5}$ GeV$^2$ accessed by any lepton scattering experiment and at the same time cover up to Q$^2$ of 6$\times$10$^{-2}$ GeV$^2$ in a single fixed experimental setup. The lowest Q$^2$ range and hence the lowest scattering angles (0.5 - 0.7 deg.) will be covered with the help of a new cross-shaped scintillator detector with a square hole in the center, placed 25 cm downstream of the target. The projected $\sim$0.43\% total uncertainty and the enhanced Q$^2$ coverage of PRad-II will enable us to access the lowest Q$^2$ range reached in lepton scattering experiments, thereby enhancing the robustness of the extracted charge radius and help establish a new precision frontier in electron scattering. It will also help address the difference between the results from PRad and all modern electron scattering experiments, in particular Mainz 2010 $-$ the most precise electron scattering measurement to date. Finally, as the most precise lepton scattering experiment, PRad-II will examine possible systematic differences between the $e-p$ and $\mu$H results. 

\section{Acknowledgements}
This work was funded in part by the U. S. National Science Foundation (NSF MRI PHY-1229153) and by the U.S. Department of Energy (Contract No. DE-FG02-03ER41231), including contract No. DE-AC05-06OR23177 under which Jefferson Science Associates, LLC operates Thomas Jefferson National Accelerator Facility. We are also grateful to all granting agencies for providing funding support to the authors throughout this project.

\end{document}